\documentclass[journal]{IEEEtran}
\usepackage{listings}
\usepackage{xcolor}
\colorlet{punct}{red!60!black}
\definecolor{background}{HTML}{EEEEEE}
\definecolor{delim}{RGB}{20,105,176}
\colorlet{numb}{magenta!60!black}
\lstdefinelanguage{json}{
    basicstyle=\normalfont\ttfamily,
    numbers=left,
    numberstyle=\scriptsize,
    stepnumber=1,
    numbersep=2pt,
    showstringspaces=false,
    breaklines=true,
    frame=lines,
    backgroundcolor=\color{background},
    literate=
      {:}{{{\color{punct}{:}}}}{1}
      {,}{{{\color{punct}{,}}}}{1}
      {\{}{{{\color{delim}{\{}}}}{1}
      {\}}{{{\color{delim}{\}}}}}{1}
      {[}{{{\color{delim}{[}}}}{1}
      {]}{{{\color{delim}{]}}}}{1},
}

\usepackage{multirow}
\usepackage{cite}

\ifCLASSINFOpdf
   \usepackage[pdftex]{graphicx}
   \DeclareGraphicsExtensions{.pdf,.jpeg,.png}
\else
\fi
\usepackage[cmex10]{amsmath}
\usepackage{amsthm}
\usepackage{amssymb}
\usepackage{algorithm}
\usepackage{algpseudocode}

\algnewcommand\algorithmicinput{\textbf{Parameter:}}
\algnewcommand\Param{\item[\algorithmicinput]}

\usepackage{array}
\usepackage{caption}
\usepackage{subcaption}

\usepackage{fixltx2e}
\usepackage{url}

\usepackage{balance}

\newtheorem{theorem}{Theorem}
\newtheorem{proposition}{Proposition}
\newtheorem{corollary}{Corollary}
\newtheorem{lemma}{Lemma}[theorem]
\newcommand*{\ROOT}{.}

\newenvironment{alphafootnotes}
{\par\edef\savedfootnotenumber{\number\value{footnote}}
	
	\setcounter{footnote}{0}}
{\par\setcounter{footnote}{\savedfootnotenumber}}

\begin{document}
%
\title{Feature-Sharing in Cascade Detection Systems with Multiple Applications}



%
%
%

\author{Long~N.~Le,~\IEEEmembership{Student Member,~IEEE,}
        and~Douglas~L.~Jones,~\IEEEmembership{Fellow,~IEEE}
\thanks{L.N. Le and D.L. Jones are with the Department
of Electrical and Computer Engineering, University of Illinois at Urbana-Champaign, Illinois,
IL, 61801 USA.  D.L. Jones is currently Director of the Advanced Digital Sciences Center.}
}

%
%


\maketitle

\begin{abstract}
Traditional distributed detection systems are often designed for a single target application. However, with the emergence of the Internet of Things (IoT) paradigm, next-generation systems are expected to be a shared infrastructure for multiple applications. To this end, we propose a modular, cascade design for resource-efficient, multi-task detection systems. Two (classes of) applications are considered in the system, a primary and a secondary one. The primary application has universal features that can be shared with other applications, to reduce the overall feature extraction cost, while the secondary application does not. In this setting, the two applications can collaborate via feature sharing. We provide a method to optimize the operation of the multi-application cascade system based on an accurate resource consumption model. In addition, the inherent uncertainties in feature models are articulated and taken into account.
For evaluation, the twin-comparison argument is invoked, and it is shown that, with the optimal feature sharing strategy, a system can achieve 9$\times$ resource saving and 1.43$\times$ improvement in detection performance.

\end{abstract}



%
\IEEEpeerreviewmaketitle


\section{Introduction}\label{sec:intro}

Traditional distributed detection systems are often designed for a single target application. However, with the emergence of the Internet of Things (IoT) paradigm, next-generation systems are expected to be a shared infrastructure for multiple applications, and hence require rethinking of the overall system design.

To support multiple tasks seamlessly, a detection system needs to be modularly designed, i.e.\ made up of components that are reusable for various applications with different objectives and constraints. A similar view is shared by the TerraSwarm Research Center \cite{lee2012terraswarm}, whose aim is to create software components that serve as building blocks for IoT application developers. Likewise, Atzori et al.\ \cite{atzori2010internet} proposed a service-oriented architecture for the IoT, where an ecosystem of services lays the foundation for IoT applications to be built on top.

Like many system design problems, resource-efficiency is an important objective in the design of multi-task detection systems \cite{le2013energy}. A well-known  resource-efficient, modular design is the cascade structure, which consists of a collection of detection modules in tandem. The design has been applied successfully in the single-application context, e.g.\ face detection \cite{viola2001rapid}. However, we propose that the cascade also has a great potential in the multiple-application context. Namely, thanks to its modular design, modules in the cascade could be shared between applications. In addition, the output of a module can be used to dynamically guide/control the execution of others in the system, providing necessary degrees of freedom to optimize the trade-off between system resource consumption and detection performance.

\begin{figure}
	\centering
	\includegraphics[width=\linewidth]{\ROOT/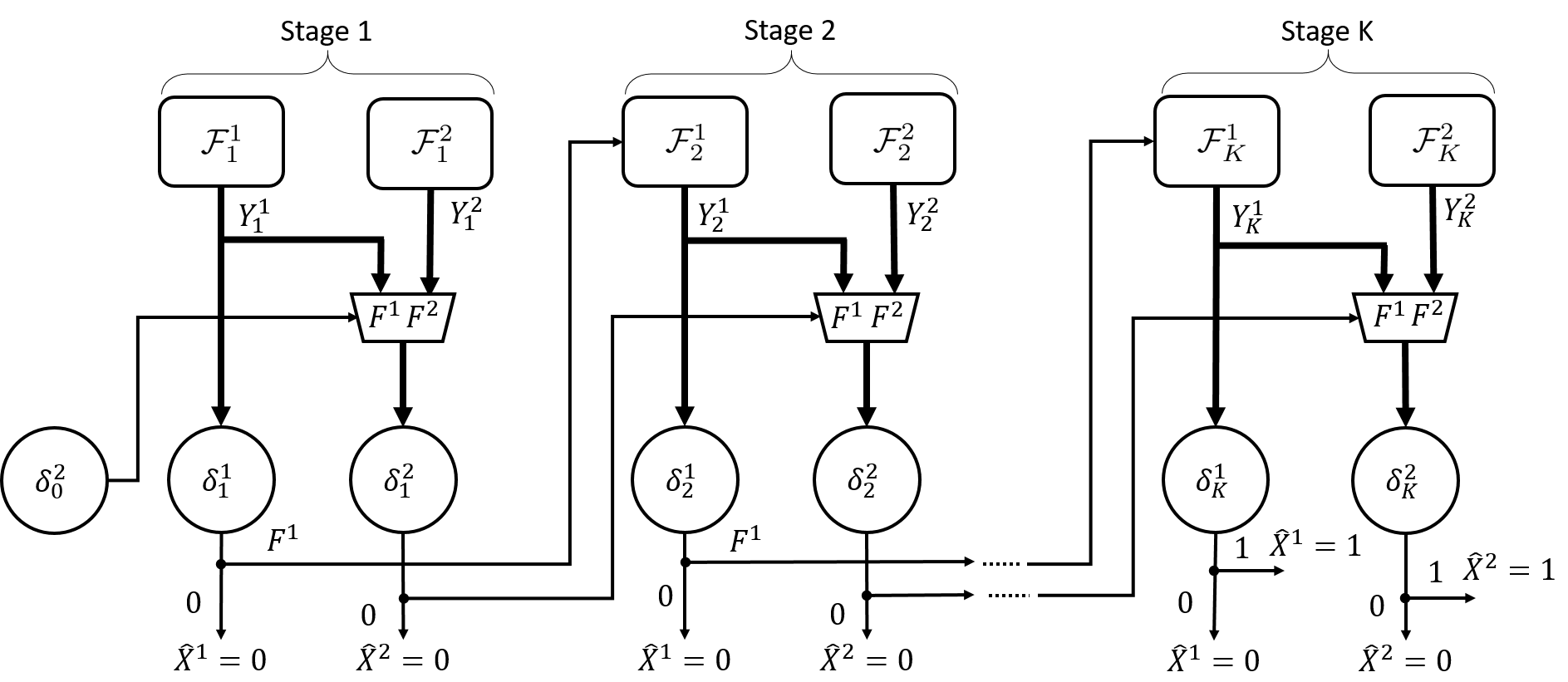}
	\caption{The cascade detection system with 2 applications (indexed by superscripts) and $K$ stages/layers (indexed by subscripts). For stage $i$ of application $j$, $\mathcal{F}_i^j$ denotes the feature extractor and $\delta_i^j$ denotes the binary decision of a detector. The feature itself is denoted by $Y_i^j$. $X^j$ is the (detection) target state, and $\hat{X^j}$ is the prediction about $X^j$ by a detector.}
	\label{fig:cascade}
\end{figure}

In this paper, we specifically study the multi-application cascade detection system whose model is shown in Fig.\ \ref{fig:cascade}. It is assumed that there are two applications and $K$ layers in the system. The system is illustrated in Fig.\ \ref{fig:cascade}, where the superscripts are used to index applications, and the subscripts are used to index stages.

Each layer of the cascade is occupied by detection modules/detectors from both applications. Ignoring the application index (superscript) for now, a detector at stage $i$ consists of a feature extractor $\mathcal{F}_i$, which produces the feature $Y_i$, and a decision rule $\delta_i$, which takes $Y_i$ and all previous features $Y_1,\dots,Y_{i-1}$ as input. $\delta_i$ outputs different values depending on both the application and the stage (see Eq.\ \eqref{eqn:delta1} and \eqref{eqn:delta2}). $X$ is the state of the (detection) target, which takes value $1$ when the target is present, and $0$ otherwise. Finally, $\hat{X}$ denotes the prediction of $X$ by the detector.

Of the two applications, we let one be the \textit{primary} and the other be the \textit{secondary}, denoted by superscript indices $1$ and $2$ in Fig.\ \ref{fig:cascade}, respectively. The primary application is the one whose feature extractors produce \textit{universal features} that are sharable. In contrast, features produced by the secondary application are not universal and hence assumed to produce no value in sharing. This distinction is illustrated in Fig.\ \ref{fig:cascade}, where primary features $Y_i^1, i=1,\dots,K$ can be shared with the secondary application, but secondary features $Y_i^2, i=1,\dots,K$ can only be used by the secondary application. Examples of universal features for audio applications are signal energy, the Mel-frequency cepstrum coefficient (MFCC) \cite{salamon2014dataset} and time-varying sinusoidal features \cite{mcaulay1986speech}, which have been used extensively in various audio/speech inference applications. Examples of secondary features are internal representation in neural networks, such as that of autoencoders. It is worth noting that the definition of primary and secondary applications here has no relevance to the practical importance of each, and the two-application assumption is meant for simplicity, without loss of generality. In fact, our result can be easily generalized to an arbitrary number of applications in each class (primary,secondary).

The decisions of the primary application $\delta_i^1$ can take on the following values depending on the layer/stage.
\begin{equation}\label{eqn:delta1}
\begin{aligned}
\delta_i^1 &= 
\begin{cases}
0: \text{ stop and declare } \hat{X}^1=0 \text{ (negative)}\\
F^1: \text{ use the primary feature next}\\
\end{cases}\\
&i = 1,\dots,K-1\\
\delta_K^1 &= 
\begin{cases}
0: \text{ declare } \hat{X}^1 = 0 \text{ (negative)}\\
1: \text{ declare }\hat{X}^1 = 1 \text{ (positive)}
\end{cases}
\end{aligned}
\end{equation}
Note that only negative decisions, i.e.\ $\hat{X} = 0$, are allowed at intermediate stages ($i =1,\dots,K-1$) since the goal is not to make the final decision (which is reserved for the last stage with the best performance) but to screen out early negative instances, which is more likely in a rare-target setting.

Since the secondary application has access to both primary and secondary features, its decisions $\delta_i^2$ at intermediate stages has more options and are given as follows.
\begin{equation}\label{eqn:delta2}
\begin{aligned}
\delta_0^2 &= 
\begin{cases}
F^1: \text{ use the primary feature next}\\
F^2: \text{ use the secondary feature next}\\
\end{cases}\\
\delta_i^2 &= 
\begin{cases}
0: \text{ stop and declare } \hat{X}^2=0 \text{ (negative)}\\
F^1: \text{ use the primary feature next}\\
F^2: \text{ use the secondary feature next}\\
\end{cases}\\
&i = 1,\dots,K-1\\
\delta_K^2 &= 
\begin{cases}
0: \text{ declare }\hat{X}^2 = 0 \text{ (negative)}\\
1: \text{ declare }\hat{X}^2 = 1 \text{ (positive)}
\end{cases}
\end{aligned}
\end{equation}
Namely, intermediate decisions include feature selection and (early) negative decision making. Note that $\delta_0^2$ is the decision occurs before any features are observed, and thus restricted from making early negative decisions.


The cascade structure has been studied before in the literature. For instance, the seminal work by Viola and Jones \cite{viola2001rapid} showed empirically that such a design is very effective in detecting rare targets in a large dataset, and was also proposed as a resource-efficient approach for stream mining by Turaga et al.\ in \cite{turaga2006resource}. However, existing works either 1) offer solutions that have limitations, to be articulated in Section \ref{sec:related}, or 2) focus on a single application. Our study here involves the cascade structure with multiple applications, investigates the potential of sharing features between them, and offers a solution that does not have limitations of prior works. For instance, our resource-consumption model is more accurate than existing works, which often suffers from a `nebulous' resource-consumption model that is inapplicable in practice. Furthermore, it is observed that there are inherent uncertainties in some features of the cascade, and an approach is proposed to address them. Beside optimizing parameters of the cascade, we also show that, under mild conditions, the cascade design itself is optimal. That is, adding additional degrees of freedom such as early positive decisions to the cascade structure does not improve its performance. 

The rest of the paper is organized as follows. Section \ref{sec:related} reviews prior works that studied the cascade structure, along with their limitations. Section \ref{subsec:featModel} discusses feature models and their potential uncertainties, then Section \ref{subsec:sysOpt} presents our formulation and solution for the multi-application cascade system. A system simulation and final remarks are given in Sections \ref{sec:sim} and \ref{sec:concl}, respectively.

\section{Prior works}\label{sec:related}
It is worthwhile to note that the cascade detection system of interest here is different from the serial detector network in the distributed detection literature \cite{tang1991optimization,swaszek1993performance,viswanathan1988optimal}, in which the decision of a current module is treated as an extra observation, instead of as a control signal to \textit{censor} subsequent modules and conserve resources.

The cascade architecture is prevalent in many inference applications, with the most widely-known example being the seminal work in face detection by Viola and Jones \cite{viola2001rapid}. In \cite{viola2001rapid}, the system of cascaded detection modules is used to quickly discard many negative sub-images typically observed in face-detection applications, thus significantly speeding up the detection process. However, the cascade is not optimized in \cite{viola2001rapid}, leaving the optimal classifiers' parameters, both thresholds and weights, to be desired.

To this end, Luo \cite{luo2005optimization} proposed to optimize thresholds of each detection module in a cascade using the classical Neyman-Pearson criterion, without consideration of resource cost. Under the assumption of statistical independence between detection modules, a gradient-based algorithm is proposed to search for the locally optimal thresholds, which is also a limitation of \cite{luo2005optimization}. In contrast, our approach guarantees a globally optimal, resource-aware solution and does not assume independence between stages.

Later, Jun and Jones \cite{jun2010energy} incorporated an energy resource constraint in the Neyman-Pearson-based optimization over thresholds of a two-stage cascade. In this setting, three solution types were identified: one that utilizes all of the available energy and false-alarm rate, one that utilizes all the energy while slacking the false-alarm constraint, and one that utilizes all the false alarm while slacking the energy constraint. An algorithm to find the optimal thresholds is only available if the true solution is of the first type. Later, it is proven in \cite{jun2013cascading} that, if observations of the first stage are reused in the second stage, then the first-type solution is optimal. In contrast, our approach generalizes to an arbitrary number of stages.

Chen et al.\ \cite{chen2012fidelity} designed a surveillance system using a two-stage cascade of low-end (acoustic and infrared) and high-quality (camera) sensors. The system in \cite{chen2012fidelity} can find a triggering threshold that either minimizes the detection error, or satisfies a constraint on the CPU utilization for video processing, but not both, and a heuristic was used to combine the two solutions, i.e.\ use the threshold that minimizes the detection error if it also satisfies the utilization constraint, otherwise use the one that satisfies the constraint. Unlike the ad-hoc approach of \cite{chen2012fidelity}, our solution is derived from a well-defined framework.
It is worth noting that Cohen et al.\ \cite{cohen2013managing} also studied a similar problem in which a multi-modal sensing system (with a PIR sensor and a camera) was designed for monitoring vehicles. While the treatment in \cite{cohen2013managing} is principled (based on the partially-observable Markov decision process (POMDP) framework), the sensors are \textit{not} operated in cascade, but instead are equally plausible options at each time step, and hence is different from our work.

Since the optimization of the cascade is hard,
Raykar et al.\ \cite{raykar2010designing} relaxed the problem by 
assuming classifiers in the cascade produce soft/probabilistic 
outputs instead of hard decisions, and converted the joint optimization of
classifiers' linear weights into a maximum {\it a posteriori} problem.
Feature costs are also incorporated into the optimization using the
standard Lagrangian argument, and a gradient-based algorithm is used to
find the optimal weights. However, the thresholds must be found using
an exhaustive grid search, which is computationally intensive
for cascades with many classifiers. Our solution does not suffer this drawback.

Chen et al.\ \cite{chen2012classifier} proposed a cyclic optimization algorithm to optimize the linear weights of the classifiers in the cascade, along with their early-exit thresholds. That is, at each iteration, the algorithm cycles through all classifiers in the cascade, optimizing each one while leaving others untouched. The algorithm stops when the loss function no longer improves. A disadvantage of such optimization procedure is that it requires multiple passes through the cascade, and there is no theoretical bound on the number of iterations it will take. In contrast, our solution requires only a single pass through the cascade.

In stream mining, Turaga et al.\ \cite{turaga2006resource} employed a cascade of Gaussian mixture model (GMM)-based classifiers and formulated a problem to find both the number of mixture components and the threshold in each classifier that maximize the system detection rate subject to constraints on false alarm, memory and CPU. The solution in \cite{turaga2006resource} takes a person-by-person approach where it iterates between 1) finding optimal numbers of mixture components, i.e.\ resource allocation, for all classifiers given thresholds, and 2) finding optimal thresholds for a given resource allocation. However, this approach failed to capture the direct dependence of the cascade's resource consumption on its thresholds, and is inherently suboptimal.

A limitation of the above works is that they only considered open-loop solutions where the thresholds are independent variables to be optimized. Ertin \cite{emre1999polarimetric} considered closed-loop solutions for the two-stage cascade detection problem where the optimal decision rule at each stage, which is observation-dependent, is sought. It was shown that the optimal policies are still likelihood ratio tests, but with coupling thresholds, i.e.\ the threshold at a stage depends on the receiver operating characteristic (ROC) and the threshold of the other stage. Namely, the optimal thresholds can not be found using the solution technique employed by \cite{emre1999polarimetric}. Note that, unlike classical detection problems, optimizing thresholds in a cascade is critical in the trade-off between inference performance and resource cost. A contribution of this paper is finding the optimal parameters (both test-statistics and thresholds) for general cascade detection systems.

Trapeznikov et al.\ studied a generalization of the cascade that was termed multi-stage sequential reject classifier (MSRC), which is simply the cascade with an additional positive decision \cite{trapeznikov2013multi} or multiple additional (classification) decisions\cite{trapeznikov2013supervised} at intermediate stages. Their resource-consumption model is `nebulous', i.e.\ if the decision at an intermediate stage is to defer to the next stage, an abstract, \textit{independent} "penalty" is incurred. In contrast, in our resource model, these penalties are shown to be precisely the Lagrangian-weighted of the feature extraction costs, and hence they are coupled (see Eq.\ \eqref{eqn:VMultiApp}).

On the other hand, a resource-consumption model closely relates to ours was considered by Wang et al.\ in \cite{wang2014lp}. The minor difference is that, instead of being proposed, our model was derived from first principle. However, \cite{wang2014lp} formulated the problem using the empirical risk minimization framework since it was assumed that probabilistic models of high-dimensional features cannot be estimated. We take a different approach where it is assumed that probabilistic models of features \textit{can} be estimated, by first reducing the features' dimensionality. In other words, the input into our algorithm are (probabilistic) models, not a dataset as in \cite{wang2014lp}. In addition, the solution proposed in \cite{wang2014lp} is a convex linear-program, which requires a convex relaxation (with an upper-bounding convex surrogate function) of the true objective function. In contrast, our solution is a dynamic program and requires no relaxation.

\section{Optimizing the multiple-application cascade system}
\label{sec:cascade}

\subsection{Feature models}\label{subsec:featModel}
The discussion in this section is applicable to both applications, and hence the application indices (superscripts) are dropped.
For the rest of the document, the colon notation is used to denote a collection, e.g.\
\begin{equation}
y_{1:i} \triangleq \{ y_1, \dots, y_{i-1}, y_i\}
\end{equation}

Recall that $Y_i$ denotes the feature used by the detector at stage $i$, and is modeled as a random variable whose distribution depends on the latent target $X \in \{0,1\}$, i.e.\
\begin{equation}\label{eqn:probModel}
\begin{aligned}
Y_i &\sim \mathrm{p}_i(y_i | x),  x \in \{0,1\}, i = 1,\dots, K
\end{aligned}
\end{equation}
where lower-case letters denote realizations of the corresponding random variable in upper case and $\mathrm{p}$ denotes a probability mass/density function. It is assumed that these distributions are stationary and hence can be estimated during training. To handle the non-stationarity case, a straightforward, yet naive method, is to perform periodic retraining. More sophisticated methods can be investigated in future work. 

Using Bayes' rule, the posterior probability of target presence is given by
\begin{equation}\label{eqn:genModel}
\begin{aligned}
&\pi_1(y_1) = \frac{1}
{ 1+\frac{1-\pi_0}{l_1(y_1)\pi_0} }\\
&\pi_i(y_{1:i}) = \frac{1}
{ 1+ \frac{1- \pi_{i-1}(y_{1:i-1})}{l_i(y_i) \pi_{i-1}(y_{1:i-1})} }\\
&\hspace{2cm}i = 2,\dots, K
\end{aligned}
\end{equation}
where $l_i(y_i) \triangleq \mathrm{p}_i(y_i|1)/\mathrm{p}_i(y_i|0)$ and $\pi_i(y_{1:i}) \triangleq \mathrm{P}(X=1|y_{1:i})$ are the likelihood function and posterior probability at stage $i$, respectively. $\pi_0 \triangleq \mathrm{P}(X=1)$ is the prior probability of the target presence. Finally, the evidence probability is given by
\begin{equation}\label{eqn:evidenceGen}
\mathrm{p}_{i}(y_{i}|y_{1:i-1}) = \mathrm{p}_{i}(y_{i}|1)\pi_{i-1} + \mathrm{p}_{i}(y_{i}|0)(1-\pi_{i-1})
\end{equation}

An important aspect of the cascade detection system is that, except for the last stage, the main goal of other stages is to quickly screen out negative instances, and not to make the final decision. Therefore features used at stages other than the last one are suboptimal for the detection task by  design, to keep the cost of their execution low. For instance, the all-band energy feature can neither characterize a bandpass target precisely, nor distinguish between a bandpass target and another bandpass interference, but can still be useful in the cascade thanks to its low cost \cite{jun2013cheap}. The sub-optimality of these early-stage features, either due to 1) the failure to discriminate the target against potential interferences, or 2) the insufficient modeling of the target, can all be modeled as \textit{uncertainty} in feature models. To this end, we employ the following \textit{least-favorable} feature density models, developed by Huber in the context of robust detection \cite{huber1968robust},\cite[Chapter 10]{huber2011robust},\cite[Chapter 6]{levy2008principles}, in place of the nominal ones.
\begin{equation}\label{eqn:leastFavor}
\begin{aligned}
\mathrm{p}_i(y|0) &\leftarrow 
\begin{cases}
\frac{1-\epsilon_{0i}}{v'+w'l_{Li}} [v'\mathrm{p}_i(y|0) + w'\mathrm{p}_i(y|1)], l_i(y)<l_{Li}\\
(1-\epsilon_{0i}) \mathrm{p}_i(y|0), l_{Li}\leq l_i(y)\leq l_{Ui}\\
\frac{1-\epsilon_{0i}}{w''+v''l_{Ui}} [w''\mathrm{p}_i(y|0) + v''\mathrm{p}_i(y|1)], l_i(y)>l_{Ui}
\end{cases} \\
\mathrm{p}_i(y|1) &\leftarrow
\begin{cases}
\frac{(1-\epsilon_{1i})l_{Li}}{v'+w'l_{Li}} [v'\mathrm{p}_i(y|0) + w'\mathrm{p}_i(y|1)], l_i(y)<l_{Li}\\
(1-\epsilon_{1i}) \mathrm{p}_i(y|1), l_{Li}\leq l_i(y)\leq l_{Ui}\\
\frac{(1-\epsilon_{1i})l_{Ui}}{w''+v''l_{Ui}} [w''\mathrm{p}_i(y|0) + v''\mathrm{p}_i(y|1)], l_i(y)>l_{Ui}
\end{cases}\\
&i = 1,\dots,K-1
\end{aligned}
\end{equation}
where the `$\leftarrow$' symbol is the assignment operator and
\begin{equation}
\begin{aligned}
v' = \frac{\epsilon_{1i}+\nu_{1i}}{1-\epsilon_{1i}}, 
v'' = \frac{\epsilon_{0i}+\nu_{0i}}{1-\epsilon_{0i}}\\
w' = \frac{\nu_{0i}}{1-\epsilon_{0i}}, 
w'' = \frac{\nu_{1i}}{1-\epsilon_{1i}}
\end{aligned}
\end{equation}
and $0 \leq\epsilon_{0i},\epsilon_{1i},\nu_{0i},\nu_{1i} \leq 1$ are uncertainty parameters of stage $i$. $l_{Li}$ and  $l_{Ui}$ are the lower and upper bounds of the likelihood ratio at stage $i$, respectively, and can be found by solving the equations outlined in \cite[Chapter 6]{levy2008principles}. Note that since the new least-favorable densities result in a bounded likelihood function, the corresponding posterior probability is also bounded.
\begin{equation}\label{eqn:piRobust}
\begin{aligned}
\pi_{Li} \triangleq \frac{1}{1+\frac{1-\pi_{i-1}}{l_{Li}\pi_{i-1}}} \leq \pi_i(y_{1:i}) \leq 
\pi_{Ui} \triangleq \frac{1}{1+\frac{1-\pi_{i-1}}{l_{Ui}\pi_{i-1}}}
\end{aligned}
\end{equation}




\subsection{System model and optimization}\label{subsec:sysOpt}
Optimizing the cascade system amounts to finding optimal decision rules $\delta_{1:K}^{1:2}$ that jointly minimize the proposed system's Bayes risk $R_B$ of incorrect decisions subject to an expected system resource (e.g.\ energy) constraint $e$.
\begin{equation}\label{eqn:mainObj}
\begin{aligned}
&\min_{\delta_{1:K}^{1:2}} R_B(\delta_{1:K}^{1:2}) \triangleq \sum_{j=1}^2 R_B^j(\delta_{1:K}^j)\\
&s.t.\  E(\delta_{1:K}^{1:2}) \triangleq \sum_{j=1}^2 E^j(\delta_{1:K}^j) \leq e
\end{aligned}
\end{equation}
where $E$ is the expected system resource consumption. It is assumed that the total Bayes risk and expected resource consumption of the system is the sum from those of the individual application, i.e.\ $R_B^j$ and $E^j$, $j=1,2$. The Lagrangian technique can be used to convert the constrained optimization problem \eqref{eqn:mainObj} into the following unconstrained, but regularized, one
\begin{equation}\label{eqn:minR}
\begin{aligned}
\min_{\delta_{1:K}^{1:2}} R(\delta_{1:K}^{1:2}) &\triangleq \sum_{j=1}^{2} R^j(\delta_{1:K}^j) \\
&\triangleq \sum_{j=1}^{2} (\lambda E^j + R_{K,\text{A}}^j + \sum_{i=1}^{K}R_{i,\text{M}}^j)
\end{aligned}
\end{equation}
where the parameter $\lambda$, which depends on the resource constraint $e$, couples the resource consumptions of all stages together and $R$ denotes the \textit{system risk} (with $R^j$ denotes the risk of application $j$), which is a measure of the combined detection performance and resource consumption. The Bayes risk $R_B^j$ has been broken down into multiple terms. For application $j$, $R_{i,M}^j,i=1,\dots,K-1$ are the miss (false negative) risks due to early negative decisions at intermediate stages. $R_{K,M}^j,R_{K,A}^j$ are the miss and false-alarm (false positive) risks due to incorrect decisions at the last stage. Note that the system has no false-alarm risk at intermediate stages, since the cascade structure does not allow early positive decisions to be made. Such a constraint is useful to reduce the false positive decision rate especially under model uncertainty and the target is rare.

For application $j$, the expected resource consumption at stage $i$ is the resource cost of feature extraction, denoted by $D_i^j$, weighted by the probability of the feature being selected by the previous stage. Hence,
\begin{equation}\label{eqn:lambdaE}
\begin{aligned}
E^1 &\triangleq D_1^1 + \sum_{i=1}^{K-1} D_{i+1}^1\mathrm{P}(\delta_i^1 = F^1) \\ 
E^2 &\triangleq\sum_{i=0}^{K-1} D_{i+1}^2\mathrm{P}(\delta_i^2 = F^2)\\
\end{aligned}
\end{equation}
where $D_1^1$ is weighted by $1$ because the first-stage primary feature is always extracted. Note that $D_i^j$ can be measured in practice by profiling the feature-extraction process.

The solution to Problem \eqref{eqn:minR} is given by the following theorem.

\begin{theorem}\label{thm:optDecRules}
	(The optimal decision rules for all applications in the cascade)
	For the primary application,
	\begin{equation}\label{eqn:optDecRules}
	\begin{aligned}
	\delta_i^{1\ast}(\pi_i^1) &= 
	\begin{cases}
	0, \pi_i^1(y_{1:i}^1) < \tau_i^{1\ast}\\
	F^1, \text{ else}
	\end{cases}\\
	&i = 1,\dots,K-1\\
	\delta_K^{1\ast}(\pi_K^1) &= 
	\begin{cases}
	0, \pi_K^1(y_{1:K}^1)  < \tau_K^{1\ast}\\
	1,  \text{ else}
	\end{cases}\\
	\end{aligned}
	\end{equation}
	
	For the secondary application,
	\begin{equation}\label{eqn:optDecRules2}
	\begin{aligned}
	\delta_0^{2\ast}(\pi_0^2;\pi_0^1) &= 
	F^1, \forall\pi_0^2,\forall\pi_0^1\\
	\delta_i^{2\ast}(\pi_i^2;\pi_i^1) &= 
	\begin{cases}
	0, \pi_i^2(y_{1:i}^2) < \tau_i^{2\ast}, \pi_i^1 < \tau_i^{1\ast}\\
	F^2, \pi_i^2(y_{1:i}^2) \geq \tau_i^{2\ast}, \pi_i^1 < \tau_i^{1\ast}\\
	0, \pi_i^2(y_{1:i}^2) < \eta_i^{2\ast}, \pi_i^1 \geq \tau_i^{1\ast}\\
	F^1, \pi_i^2(y_{1:i}^2) \geq \eta_i^{2\ast}, \pi_i^1 \geq \tau_i^{1\ast}
	\end{cases}\\
	&i = 1,\dots,K-1\\
	\delta_K^{2\ast}(\pi_K^2) &= 
	\begin{cases}
	0, \pi_K^2(y_{1:K}^2)  < \tau_K^{2\ast}\\
	1,  \text{ else}
	\end{cases}
	\end{aligned}
	\end{equation}
	where $\tau_i^{j\ast}, \eta_i^{j\ast} \in [\pi_{Li}^j,\pi_{Ui}^j]$ are the optimal thresholds at stage $i$ of application $j$, provided that 
	\begin{equation}\label{eqn:condForShare}
	\begin{aligned}
	C_M^2\left\{ \mathbb{E}[\pi_i^2(Y_i^1)]-\mathbb{E}[\pi_i^2(Y_i^2)] \right\} \leq \lambda D_{i}^2 ,i = 1,\dots,K
	\end{aligned}
	\end{equation}
	with $C_M^2$ defined in Corollary \ref{corol:perf}.
	
	\begin{proof}
		See Appendix \ref{subsec:app1}
	\end{proof}
\end{theorem}

Eq.\ \eqref{eqn:optDecRules} in Theorem \ref{thm:optDecRules} shows that, for the primary application, the posterior probabilities of intermediate stages can be used to guide the execution of subsequent stages by thresholding them to decide whether to stop or extract more primary features in the next stage.
The final stage has a standard detection rule, with the posterior probability being thresholded to make a prediction about the target state.

Furthermore, Eq.\ \eqref{eqn:optDecRules2} shows that the decision rules at intermediate stages of the secondary application are not only a function of $\pi_i^2$, but are also parameterized\footnote{After the semicolon} by $\pi_i^1$. If $\pi_i^1 \geq \tau_i^{1\ast}$, then according to \eqref{eqn:optDecRules}, the next-stage primary feature is available, and the optimal decision always selects this feature over the secondary feature (feature sharing) for the next stage (as long as $\pi_i^2$ is above the threshold for early negative decision $\eta_i^{2\ast}$). Since the first-stage primary feature is always available, it is always selected by $\delta_0^{2\ast}$. Otherwise, if the primary feature will not be available because $\pi_i^1 < \tau_i^{1\ast}$, then the secondary application falls back to selecting the secondary feature, assuming $\pi_i^2$ is above the threshold for early negative decision $\tau_i^{2\ast}$. Note that the thresholds for early negative decisions are different under each case. Finally, the final stage's decision is simply a standard detection rule.

The structure of \eqref{eqn:optDecRules2} favors feature-sharing whenever possible. This policy is optimal provided that additional constraints in \eqref{eqn:condForShare} on the parameters of the cascade hold. Intuitively, \eqref{eqn:condForShare} requires that the difference between primary and secondary feature distributions is relatively small compared to the resource cost of extracting the latter. 

Finally, the optimal threshold values $\{\tau_i^{j\ast},j=1,2\}$, which are critical in this trade-off between performance and resource cost, can be found using Algorithm \ref{alg:recur_thresh}.



Given the above optimal decisions, Corollary \ref{corol:perf} quantifies the optimal performance of the multi-application cascade system.

\begin{corollary}\label{corol:perf}
	(Optimal performance of applications in the cascade)
	For the primary application,
	\begin{equation}
	R^{1\ast}(\pi_0^1) \triangleq R^1(\delta_{1:K}^{1\ast},\pi_0^1) = V_0^1(\pi_0^1)
	\end{equation}
	where $V_0^1(\pi_0^1)$ is the result of the following recursion
	\begin{equation}\label{eqn:Vrecur}
	\begin{aligned}
	V_K^1(\pi_K^1) &\triangleq \min (\underbrace{C_M^1\pi_K^1}_{\text{miss risk}}, \underbrace{C_{A}^1(1-\pi_K^1)}_{\text{false-alarm risk}}),\pi_K^1\in[0,1]\\
	V_{i}^1(\pi_{i}^1) &\triangleq \min (C_M^1\pi_i^1, \underbrace{\lambda D_{i+1}^1 + \mathbb{E}[V_{i+1}^1(\pi_{i+1}^1(Y_{i+1}^1,\pi_{i}^1))]}_{\text{expected next-stage primary value function}} ),\\
	&\pi_i^1\in [\pi_{Li}^1,\pi_{Ui}^1], i = 1,\dots, K-1\\
	V_{0}^1(\pi_{0}^1) &\triangleq \lambda D_{1}^1 + \mathbb{E}[V_{1}^1(\pi_{1}^1(Y_{1}^1,\pi_{0}^1))] \\
	\end{aligned}
	\end{equation}
	And the corresponding optimal thresholds are given by
	\begin{equation}\label{eqn:thresh}
	\begin{aligned}
	\tau_K^{1\ast} &= C_{A}^1/(C_{A}^1+C_{M}^1)\\
	\tau_i^{1\ast} &= \max(\pi_{Li}^1,\min(\pi_{Ui}^1,\min \{\pi_i^1: V_{i}^1(\pi_{i}^1) - C_M^1\pi_i^1 < 0\})),\\
	&i = 1,\dots, K-1\\
	\end{aligned}
	\end{equation}
	For the secondary application,
	\begin{equation}
	R^{2\ast}(\pi_0^2;\pi_0^1) \triangleq R^2(\delta_{1:K}^{2\ast},\pi_0^2;\pi_0^1) = V_0^2(\pi_0^2;\pi_0^1)
	\end{equation}
	where $V_0^2(\pi_0^2;\pi_0^1)$ is the result of the following recursion
	\begin{equation}\label{eqn:Vrecur2}
	\begin{aligned}
	V_K^2(\pi_K^2) &\triangleq 
	\min (C_M^2\pi_K^2, C_{A}^2(1-\pi_K^2)),\pi_K^2\in[0,1]\\
	V_{i}^2(\pi_{i}^2;\pi_{i}^1) &\triangleq 
	\begin{cases}
	\min (C_M^2\pi_i^2, \\ 
	\quad \underbrace{\lambda D_{i+1}^2 +\mathbb{E}[V_{i+1}^2(\pi_{i+1}^2(Y_{i+1}^2,\pi_{i}^2);\pi_{i}^1)])}_{\substack{ \text{expected next-stage secondary value function}\\\text{using the secondary feature} }},\\
	\quad \text{ if } \pi_i^{1} < \tau_i^{1\ast}\\
	\min (C_M^2\pi_i^2,\\ 
	\quad \underbrace{\mathbb{E}[V_{i+1}^2(\pi_{i+1}^2(Y_{i+1}^1,\pi_{i}^2);\pi_{i}^1)]}_{\substack{ \text{expected next-stage secondary value function}\\\text{using the shared primary feature} }}), \\
	\quad\text{ else}
	\end{cases}\\
	&\pi_i^2\in [\pi_{Li}^2,\pi_{Ui}^2], i = 1,\dots, K-1\\
	V_{0}^2(\pi_{0}^2;\pi_{0}^1) &\triangleq  \mathbb{E}[V_{1}^2(\pi_{1}^2(Y_{1}^1,\pi_{0}^2);\pi_{0}^1))] \\
	\end{aligned}
	\end{equation}
	and the corresponding optimal thresholds are given by
	\begin{equation}\label{eqn:thresh2}
	\begin{aligned}
	\tau_K^{2\ast} =& C_{A}^2/(C_{A}^2+C_M^2)\\
	\tau_i^{2\ast} =& \max(\pi_{Li}^2,\min(\pi_{Ui}^2,\\
	&\quad\quad\min \{\pi_i^2: V_{i}^2(\pi_{i}^2;\pi_{i}^1) - C_M^2\pi_i^2 < 0\})),\\
	&\text{ if } \pi_i^1 < \tau_i^{1\ast}\\
	\eta_i^{2\ast} =& \max(\pi_{Li}^2,\min(\pi_{Ui}^2,\\
	&\quad\quad\min \{\pi_i^2: V_{i}^2(\pi_{i}^2;\pi_{i}^1) - C_M^2\pi_i^2 < 0\})),\\
	&\text{ else },\\
	&i = 1,\dots, K-1,\\
	\end{aligned}
	\end{equation}
	where $C_{M}^j,C_{A}^j,j=1,2$ are the costs associated with miss and false-alarm decisions of application $j$. 
	
\end{corollary}

Corollary \ref{corol:perf} shows that the optimal performance achieved by each application can be found using a recursive procedure. The procedure has $K$ iterations, each corresponding to a stage in the system. Starting from the last stage $K$ and proceeding backward to $0$, the value function $V_i^j$ is recursively updated (see \eqref{eqn:Vrecur} and \eqref{eqn:Vrecur2}). The last-stage value function $V_K^j$ is the minimum of the miss and false-alarm risks across $\pi_K$. An intermediate-stage value function $V_i^j,i=1,\dots,K-1$ is the minimum of the miss risk and the \textit{expected next-stage} value function, which requires the probabilistic updates in  \eqref{eqn:genModel},\eqref{eqn:evidenceGen}.
The final value function $V_0^j$ is the minimal risk achievable by an application.

Note that the secondary application's value function at an intermediate stage is given by different expressions depending on the availability of the primary feature for the next stage, i.e.\ $\pi_i^1 \lessgtr \tau_i^{1\ast}$ (see \eqref{eqn:Vrecur2}). If the primary feature is not available for the next stage, then the expected next-stage value function is taken with respect to the secondary feature, whose extraction cost ($\lambda D_{i+1}^2$) is also included. Otherwise, the expected next-stage value function does not contain the resource cost to extract the secondary feature and the expectation is taken with respect to the primary feature.

Once a value function is known, then the corresponding optimal threshold can be found using just arithmetic operations, i.e.\ comparing the value function with the miss risk (see \eqref{eqn:thresh} and \eqref{eqn:thresh2}). For the last stage $K$, the optimal threshold can be given in closed form. Note that the intermediate stages' thresholds are capped between the upper and lower bounds due to model uncertainty (see Section \ref{subsec:featModel}). For the secondary application, depending on the availability of the primary feature, the thresholds for early negative decisions are different, and hence denoted differently ($\tau_i^2$ and $\eta_i^2$, respectively). Intuitively, if the primary threshold is low, i.e.\ the primary application consumes most of the resource budget, then the secondary application is more inclined to use the shared primary feature due to low resource budget. On the other hand, if the primary threshold is high, i.e.\ the secondary application has most of the resource budget, then the secondary feature is used more to reduce its miss and false-alarm risks.

The discussion so far has been focusing on optimizing parameters of the cascade design. A natural next question to ask is whether the constraints of the cascade design can be relaxed to further improve performance. Namely, would introducing additional degrees of freedom, i.e.\ early positive decisions in intermediate stages, to the cascade \textit{always} improve its performance? Intuitively, when model uncertainties of intermediate stages are accounted for (see Section \ref{subsec:featModel}), and it is known \textit{a priori} that the target is rare, early positive decisions are likely to have higher risk and hence are discouraged. Therefore, introducing additional early positive decisions does \textit{not} always improve the performance of the cascade. The precise conditions for which the cascade design itself is optimal is given by the following proposition.
\begin{proposition}\label{prop:vivaCascade}
	(Optimality of the cascade design)
	With model uncertainty, introducing additional early positive decisions in intermediate stages of the cascade does not improve performance, as long as
	\begin{equation}\label{eqn:condVivaCascade}
	\begin{aligned}
	\underbrace{\max\{\pi_i^j: V_i^j(\pi_i^j) - C_{A}^j(1-\pi_i^j) < 0\}}_{\text{optimal threshold for early positive decision}} > \pi_{Ui}^j,\\
	i = 1,\dots,K-1, j = 1,2
	\end{aligned}
	\end{equation}
	
	\begin{proof}
		See Appendix \ref{subsec:app2}.
	\end{proof}
\end{proposition}

The left-hand side of \eqref{eqn:condVivaCascade} is the optimal threshold corresponding to an early positive decision. Namely, these additional decisions also have threshold-based optimal policies (see Appendix \ref{subsec:app2}), and a posterior probability \textit{above} such a threshold shall trigger an early positive decision. If such a threshold is above the upper bound on the posterior probability at a stage, then its early positive decision is never selected, and hence does not affect the performance of the cascade.

\section{System simulation}\label{sec:sim}

This section applies the theory developed in Section \ref{sec:cascade} to design a multi-application, acoustic detection system.

\subsection{Hardware components}\label{subsec:hardware}

The hardware components needed for an acoustic sensing system are listed in Table \ref{tab:components}, along with their power consumption (from the datasheets). Note that these are commercially-off-the-shelf (COTS) components, without any customization. The sensor's brain (supplied at 3.6 V) is Silicon Labs's WGM110, which is a low-power wireless (wifi) chip that includes a low-power 12-bit ADC, an ARM Cortex-M3 processor, and a wifi module (among others). All the control logic and the (digital) signal processing software are assumed to be implemented on this general-purposed processor, without any ASIC\footnote{application-specific integrated circuit} or DSP\footnote{digital signal processor}. In addition, a microphone and a preamp are also a part of the acoustic sensor. The power consumption of the microphone, the preamp circuit, and the ADC, altogether is 3.6 mW and considered as the baseline of the system. Data collected by the sensor are transmitted downstream to the client, which is a ML100G-10 Next Unit of Computing (NUC) from LogicSupply. Power profiling the NUC  (using PowerBlade\cite{debruin15powerblade}) results in an average power consumption of 4.744 W (at 9 V).

\begin{alphafootnotes}
\begin{table}[H]
	\caption {Power consumption of hardware components of the acoustic sensing system.} \label{tab:components} 
	\centering
	\begin{tabular}{|c|c|}
		\hline
		\textbf{Components} & \textbf{Power consumption (mW)} \\ 
		\hline
		Electret Microphone & 0.72 \\ 
		\hline
		Preamp Circuit (OPA344) & 1.08 \\ 
		\hline
		WGM110 12-bit ADC & 1.8\footnotemark\\ 
		\hline
		WGM110 ARM core & 86.4 \\ 
		\hline
		WGM110 transmission & 900 \\
		\hline
		ML100G-10 & 4744\\
		\hline
	\end{tabular}
\end{table}
\footnotetext{From \cite{jun2013cheap}, the ADC draws 0.5 mA, which is equivalent to 1.8 mW at 3.6 V.}
\end{alphafootnotes}

\subsection{Software components}

\subsubsection{Primary application (Golden-Cheeked Warbler detection)}

The detection of the Golden-cheeked Warbler (GCW)'s (type-A) calls \cite{leonard2010variation} is considered as the primary application. Namely, $X^1 = 1$ indicates the presence of a GCW call, and $X^1 = 0$ otherwise. Since the GCW is an endangered bird species, this application has important implications for their conservation.

\begin{figure}
\centering
\includegraphics[width=\linewidth]{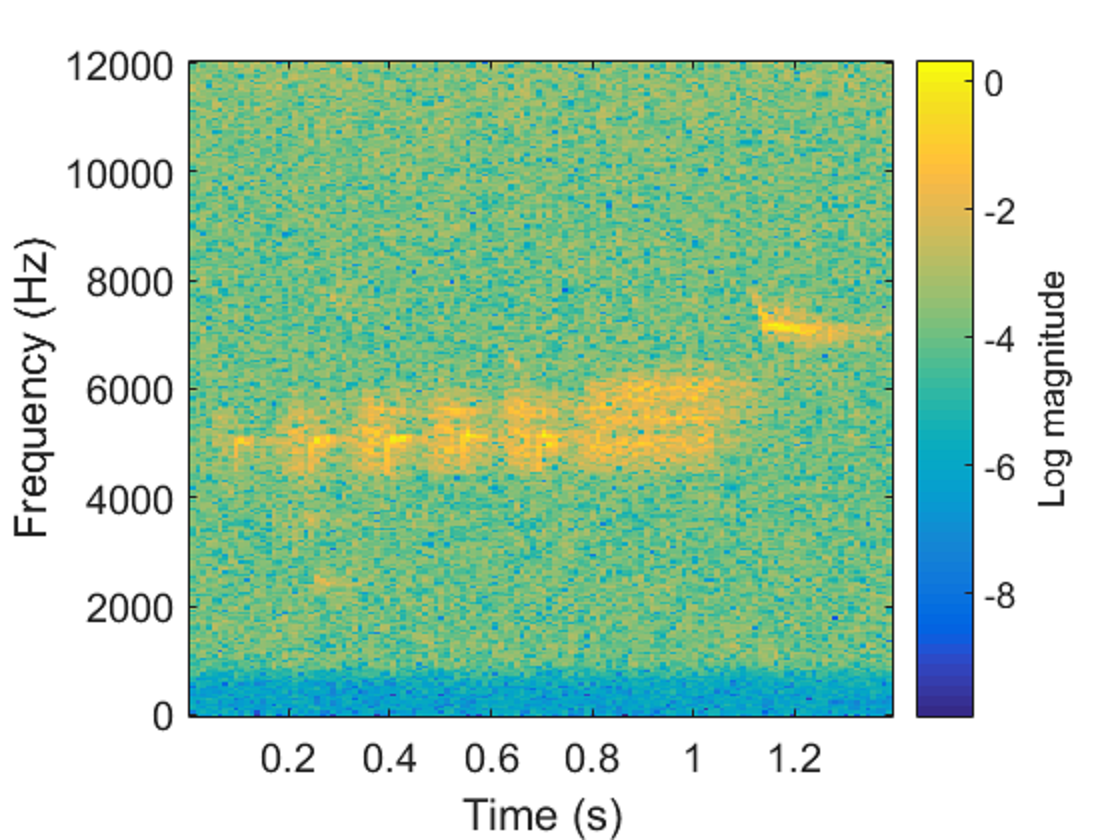}
\caption{Spectrogram of a sample GCW's (type-A) call.}
\label{fig:specgram}
\end{figure}

The application's software is organized into three subtasks: generic energy-based analysis, spectral-based analysis, and temporal-spectral-based analysis. The energy analysis is a low-complexity computation that produces energy-based features useful for detecting acoustic events from silence. The spectral-based analysis takes into account the spectral information about the GCW calls, which only has energy in the 4500-6500 Hz and 7000-8000 Hz bands (see Fig.\ \ref{fig:specgram}), to produce band-specific, energy-based features using standard DSP filtering techniques. Finally, the spectral-temporal-based analysis takes into account both the spectral and temporal structure of the GCW call from Fig.\ \ref{fig:specgram} to produce reliable, indicative features using a template matching technique. Note that the input into the above analyses is an audio stream (or precisely, its high-dimensional time-frequency representation, see Fig.\ \ref{fig:specgram}), and their output is a scalar score sequence, i.e.\ a score for each audio frame. Hence, these analyses effectively perform dimensionality reduction. 

\begin{figure}[t!]
	\centering
	\includegraphics[width=\linewidth]{\ROOT/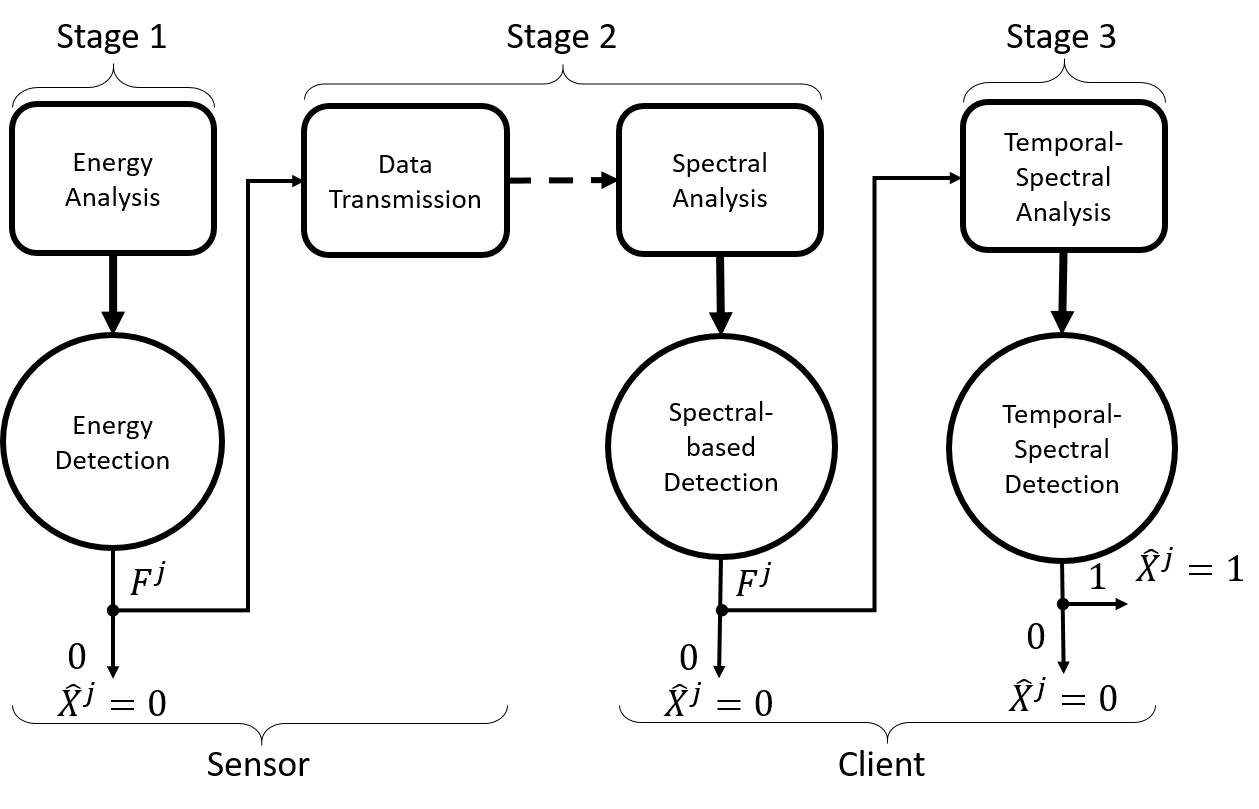}
	\caption{The software components of the primary application is organized as a cascade with 3 stages: energy analysis as stage 1, spectral-based analysis (along with the data transmission) as stage 2, and temporal-spectral analysis as stage 3. Note that components of the cascade are implemented distributedly across the network, with the dashed line representing a remote connection.}
	\label{fig:cascadeReal}
\end{figure}

Since the generic energy analysis has low computational complexity and can help prune out a significant amount of noise-only data from the audio stream early, it is executed on edge/sensor nodes. Only acoustic events are transmitted downstream to clients, where spectral and temporal-spectral-based analyses are further carried out. The system diagram is illustrated in Figure \ref{fig:cascadeReal} and arranged to fit the proposed cascade abstraction. Note that the physical separation (between sensors and clients) does not necessarily correspond to the logical separation (between stages). For instance, the cost of data transmission on sensors are included into the cost of executing the second stage, along with the cost of spectral-based analysis on clients, since they are both a result of the first-stage decision.

The resource cost parameters at each stage $D_i^1, i=1,2,3$, which can be estimated from values of Table \ref{tab:components} and the execution times of the software components, are needed to optimize the resource-performance trade-off. It is assumed that all processing finishes before a periodic deadline, i.e.\ when buffers (an ADC buffer on the sensor, a task buffer on the client) are full. The average execution time of each task (per audio frame of $32$ ms) can be estimated/profiled and is given as follows. The energy analysis takes $16$ ms\footnote{Estimated as half of the frame length.}. The average transmission time takes $11$ ms ($500$ ms for a $1.5$ s event\footnote{Profiled on a prototype.}).
Finally, the spectral and temporal-spectral analyses take $0.37$ $\mu$s and $15$ ms, respectively\footnote{Profiled in MatLab for ML100G-10.}.
Hence,
\begin{equation}
\begin{aligned}
D_1^1 &= 86.4 \times 0.016,\\
D_2^1 &= 900\times 0.011 + 4744\times 0.37\times 10^{-6}, \\
D_3^1 &= 4744\times 0.015,\\
\end{aligned}
\end{equation}

Our dataset is a 46-minute, 24 kHz audio recording at the field in Rancho Diana, San Antonio's city park.
The dataset contains 206 GCW calls (manually identified and labeled), each of whose duration is approximately one second. In addition to GCW calls, the dataset also contains various interferences from other animals' vocalization, time-varying wind noise, etc., since it is taken directly from field recording. Precisely, the fraction of GCW calls in the entire dataset is 10.19\%. Hence, this detection problem belongs to the rare-target class, where the prior is asymmetrical, i.e.\ $\pi_0^1 \ll 0.5$. In this simulation, we consider a range of prior in the rare-event regime, i.e.\ $\pi_0^1 \in [0.05,0.20]$. Finally, the miss and false-alarm costs are given by $C_M^1 = 2, C_A^1 = 1$ to emphasize that the miss risk is higher in this setting.

\begin{figure}
	\centering
	\includegraphics[width=\linewidth]{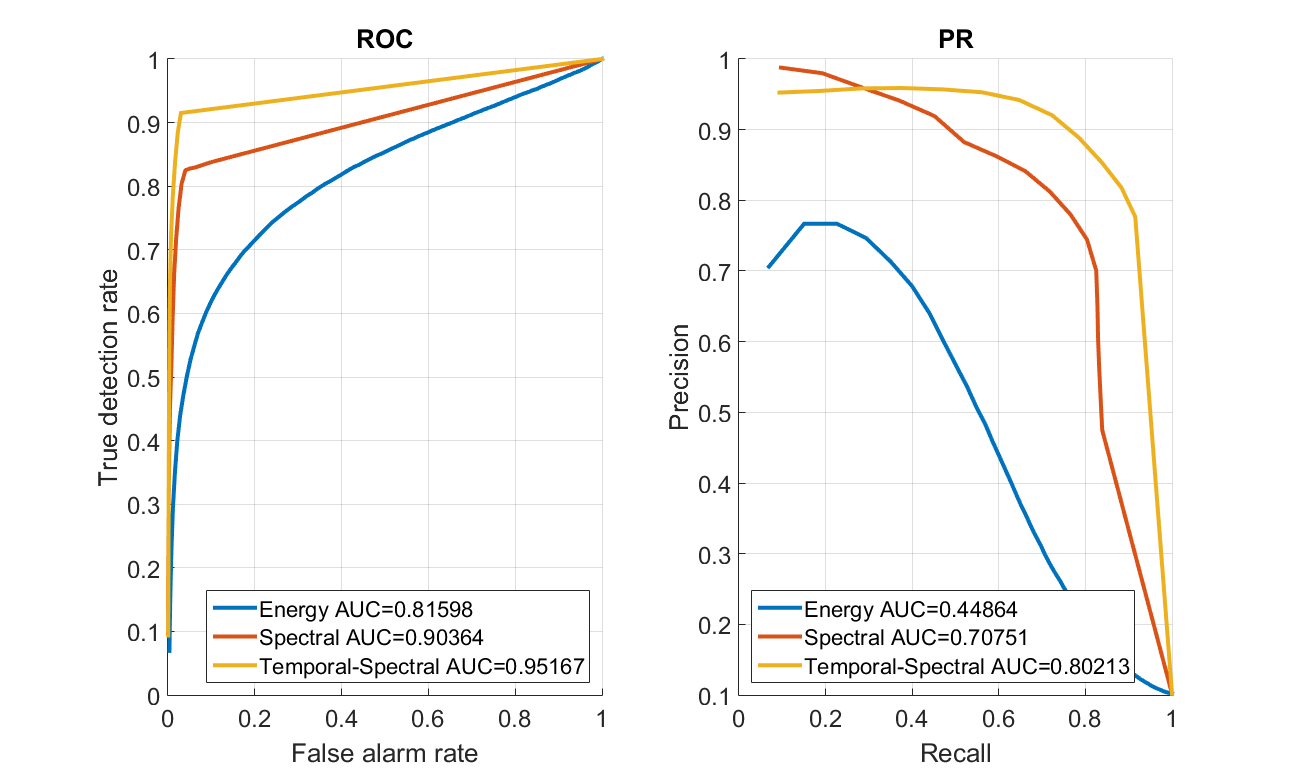}
	\caption{Receiver operating characteristic (ROC) curves and precision-recall (PR) curves of the features produced by the 3 analyses.}
	\label{fig:roc}
\end{figure}

The dataset are input to each of the three analyses discussed above. The scalar output scores from each analysis are taken as its respective features, resulting in a total of three feature sets/groups/types. The discriminative power of each feature type, or equivalently the performance of an analysis, can be quantified using receiver operating characteristic (ROC) and precision-recall (PR) curves as shown in Fig.\ \ref{fig:roc}. From the figure, it is evident that the temporal-spectral feature is better than the spectral feature, which in turn is better than the generic energy feature, at detecting GCW calls.

The conditional probability mass functions (PMF), i.e.\ $\mathrm{p}_i(y_i|x)$, of features from each analysis can be estimated up to some quantization level, i.e.\ $100$. 
Furthermore, as alluded to in Section \ref{subsec:featModel}, energy-based and spectral-based features, by construction, are inadequate to characterize GCW calls, and hence there are inherent uncertainties in these features for the detection of GCW calls. These uncertainties can be explicitly accounted for in the features' distributions using the uncertainty model discussed in Section \ref{subsec:featModel}, with the following parameters.
\begin{equation}
\begin{aligned}
\epsilon_{01}^1 &= \epsilon_{02}^1 = 0.1\\
\epsilon_{11}^1 &= \epsilon_{12}^1 = 0.1\\
\nu_{01}^1 &= \nu_{02}^1 = 0.1\\
\nu_{11}^1 &= \nu_{12}^1 = 0.1\\
\end{aligned}
\end{equation}
Intuitively, the $\epsilon$ and the $\nu$ parameters indicate the level and the strength of a contamination on the nominal distribution, respectively. A formal method to set these parameters are left for future work.
Finally, it is assumed that the temporal-spectral analysis (the last stage) is sufficient to characterize GCW calls and hence there is no uncertainty in this feature set.

\subsubsection{Secondary application}

To illustrate the benefit of feature sharing, we invoke the following twin-comparison argument. We consider a hypothetical secondary application that is, as far as the resource-performance trade-off is concerned, identical to the primary application. Namely, all parameters, such as the resource cost and the feature models at each stage, of the secondary application is the same as those of the primary one. Due to the asymmetry in feature sharing between the primary and secondary applications, it is expected that there will be differences in the resulting resource consumption and detection performance of the two applications, and the merit of the proposed feature sharing approach can be evaluated by quantifying this difference.

\subsection{Results}



The method developed in Section \ref{sec:cascade} can be used to optimize the acoustic system and the results are presented below.

\begin{figure}[t!]
	\centering
	\includegraphics[width=\linewidth]{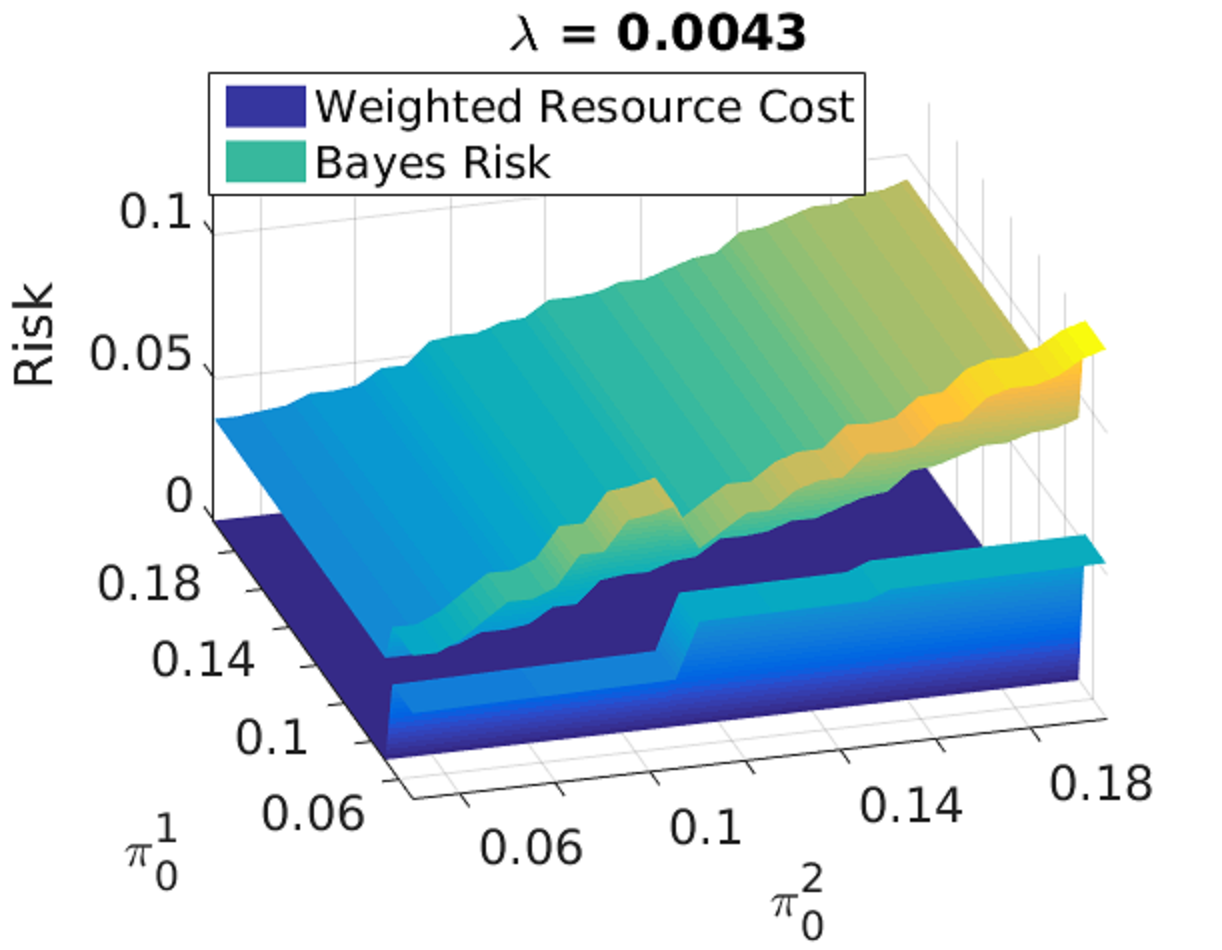}
	\caption{Breakdown of the optimal secondary risk into components (see Eq. \eqref{eqn:minR}): Detection risk and Lagrangian-weighted resource consumption. The detection risk tends to increase with the secondary prior. At a certain level, the secondary application must ramp up its resource consumption to reduce the risk.}
	\label{fig:risk2Components}
\end{figure}

Fig.\ \ref{fig:riskComponents} breaks down the primary application's risk into the weighted resource consumption, and the miss and false-alarm rates to provide an intuitive understanding of the optimal policies. Furthermore, the average resource consumption of the primary application across all priors is $44.406$ mJ (per audio frame). Without feature sharing, it is expected that the resource consumption of the secondary application would be the same as that of the primary one. However, the average resource consumption of the secondary application across the priors of interest can be found to be $4.877$ mJ (based on the break-down of the secondary risk illustrated in Fig.\ \ref{fig:risk2Components}), which is approximately a $9\times$ reduction in resource consumption. This significant resource-saving is due to the fact that extracted features are shared and not recomputed among applications. In addition, the average detection risk (both the miss and false-alarm rate) of the secondary application is also reduced by $1.43\times$ (from $4.75$\% to $3.31$\%). This reduction in risk illustrates that using shared features can be more beneficial than having no feature at all.

It is worth noting that the above applications' resource consumptions are the result of setting the regularization parameter $\lambda$ to $0.0043$, which is optimal for a resource/energy budget of $49.398$ mJ (the sum of power consumptions by both applications and the $3.6\times0.032$ mJ base line from the microphone, the preamp circuit, and the ADC of the sensor over a frame). For a given resource budget, one needs to be solved for $\lambda$. Numerical solution of the scalar variable $\lambda$ is straightforward and hence shall be skipped here.

\begin{figure}[t!]
	\centering
	\includegraphics[width=\linewidth]{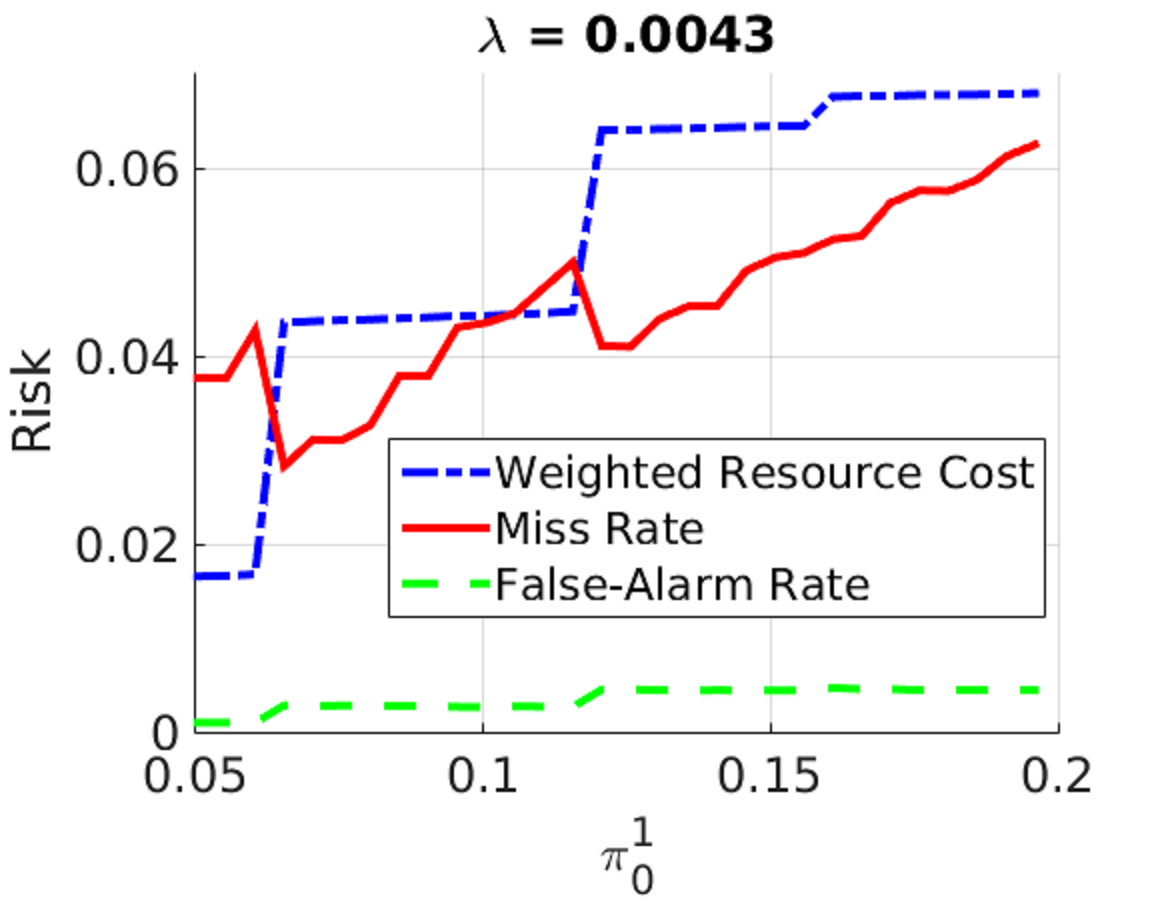}
	\caption{Breakdown of the optimal primary risk into components (see Eq. \eqref{eqn:minR}): false negative (miss), false positive (false-alarm), and Lagrangian-weighted resource consumption. Low false-alarm rate is achieved across the priors of interest. The miss rate tends to increase with the prior. At a certain level, the primary application must ramp up its resource consumption or incur more false-alarm to reduce the miss rate.}
	\label{fig:riskComponents}
\end{figure}

%
%
%
%
%

\begin{figure}[t!]
	\centering
	\includegraphics[width=\linewidth]{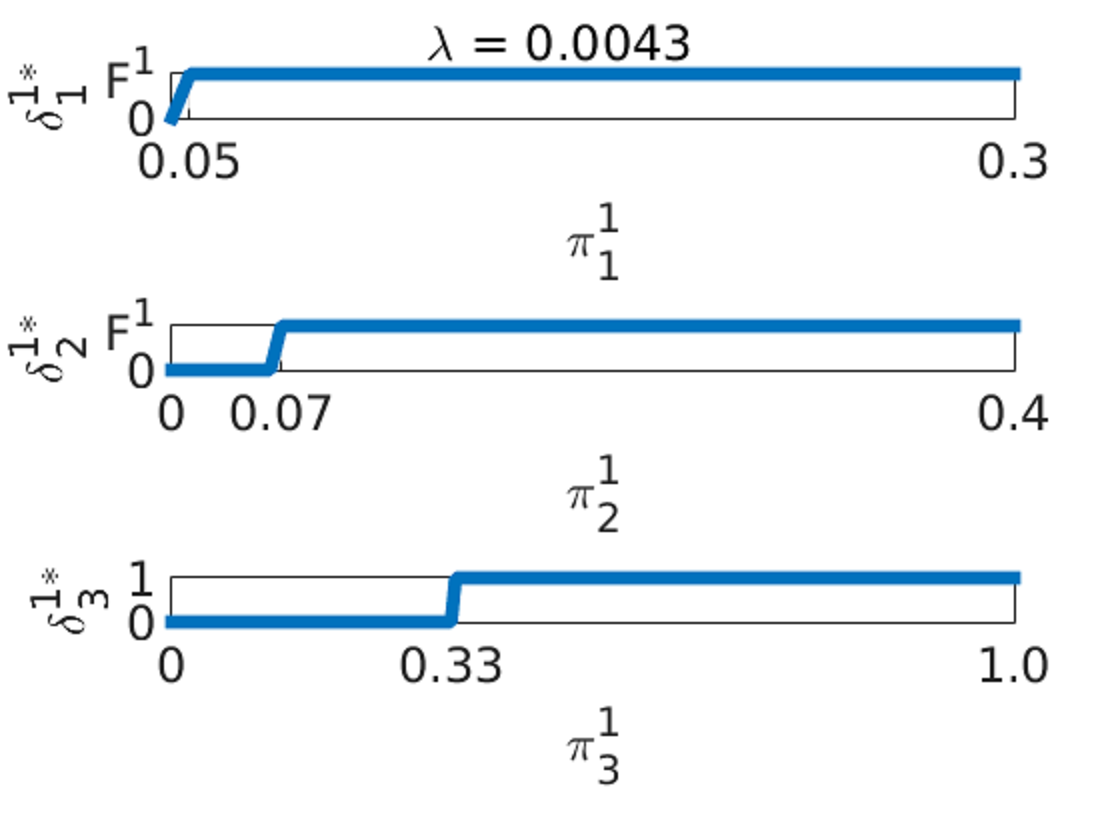}
	\caption{Optimal decision rules of the primary application $\delta_i^{1\ast}(\pi_i^{1}) \in \{F^1,0,1\}, i = 1,\dots,3$.}
	\label{fig:threshMulti}
\end{figure}

\begin{figure}[t!]
	\centering
	\includegraphics[width=\linewidth]{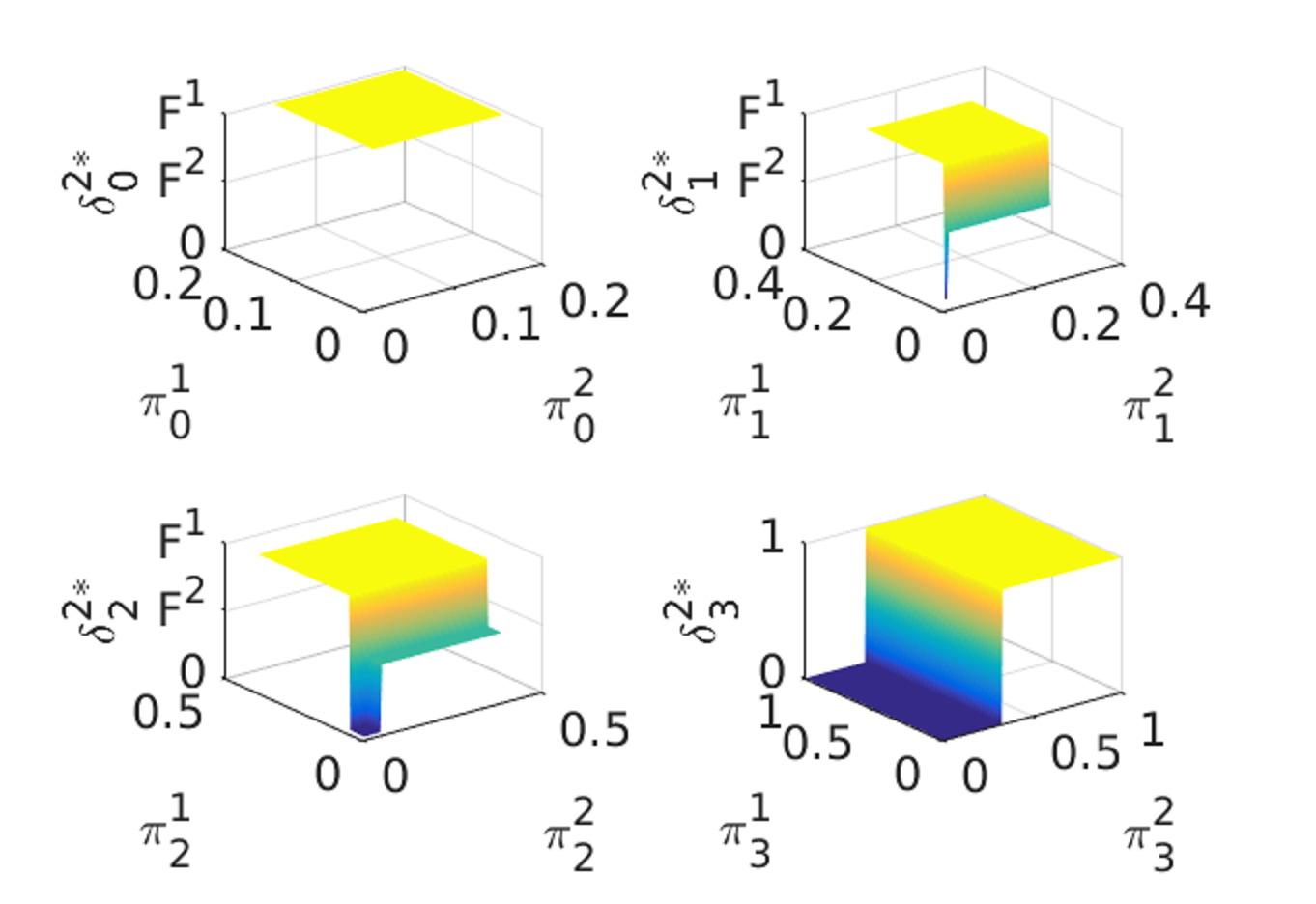}
	\caption{Optimal decision rules of the secondary application $\delta_i^{2\ast}(\pi_i^{1:2}) \in \{F^1,F^2,0,1\}, i = 0,\dots,3$.}
	\label{fig:threshMulti2}
\end{figure}

The primary and secondary decision rules are illustrated in Fig.\ \ref{fig:threshMulti} and \ref{fig:threshMulti2}, respectively. Note that the optimal policies of the secondary application take advantage of feature sharing whenever possible. Namely, $\delta_i^{2\ast} = F^1$ for all $\pi_i^2$ and $\pi_i^{1} \geq \tau_i^{1\ast}$, $i=0,1,2$. When the primary feature is not available, the secondary policies choose between extracting the secondary feature ($\delta_i^{2\ast} = F^2$) or making an early negative decision ($\delta_i^{2\ast} = 0$).

\section{Conclusion}\label{sec:concl}

This paper investigates and shows the benefits of features sharing in the optimization of resource-performance trade-off for detection systems with multiple applications.
The proposed system model focuses on the vertical design, rather than the horizontal one commonly seen in the wireless sensor network literature. Namely, it is assumed that there is only one device at each layer/stage in the system stack, as opposed to having multiple devices at the same layer. Therefore, this work can complement prior works and vice versa. For instance, Chamberland et al.\ showed that it is asymptotically optimal for all sensors in a power-constrained sensor network to adopt the same (transmission) strategy, as the number of sensors in the network goes to infinity \cite{chamberland2004asymptotic}. The result in \cite{chamberland2004asymptotic} therefore allows ours to be applicable for layers with more than one device.
Finally, a proof-of-concept implementation of the acoustic system in Section \ref{sec:sim} (with the sensor as an Android app) is available online for demonstration\footnote{At \url{http://acoustic.ifp.illinois.edu}}.

\appendices


\section*{Acknowledgements}
This work was supported in part by TerraSwarm, one of six centers of STARnet, a Semiconductor Research Corporation program sponsored by MARCO and DARPA,
and in part by a research grant for the Human-Centered Cyberphysical Systems Programme at the Advanced Digital Sciences Center from Singapore's Agency for Science, Technology and Research (A*STAR)


\appendix

\subsection{Proof of Theorem \ref{thm:optDecRules}}\label{subsec:app1}

We start by expanding the risk terms in \eqref{eqn:minR}. The false negative (miss) rate due to early negative decision for the first stage is 
\begin{equation}\label{eqn:multiR1}
\begin{aligned}
R_{1,M}^1 &= \int \mathrm{p}(\mathrm{d}y_1^{1})\big\{
C_M^1\pi_1^1(y_1^{1})\mathbb{I}(\delta_1^1 = 0)
\big\}\\
R_{1,M}^2 &= \int \mathrm{p}(\mathrm{d}y_1^{1:2})\big\{
C_M^2\pi_1^2(y_1^{2})\mathbb{I}(\delta_1^{2} = 0,\delta_0^{2} = F^2)+\\
&C_M^2\pi_1^2(y_1^{1})\mathbb{I}(\delta_1^{2} = 0,\delta_0^{2} = F^1)
\big\}\\
\end{aligned}
\end{equation}
where $R_{1,M}^1, R_{1,M}^2$ are the first-stage miss risk of the primary and secondary applications, respectively. Furthermore, the first term of $R_{1,M}^2$ is due to using the secondary feature $y_1^2$ ($\delta_0^2 = F^2$), and the second term is due to using the shared (primary) feature $y_1^1$ ($\delta_0^2 = F^1$).
$\mathbb{I}()$ denotes the indicator function that takes value $1$ when its argument (a probability event) is true and $0$ otherwise. Finally, $\mathrm{p}(\mathrm{d}y_{1:K})$ is the probability measure of feature realizations $y_{1:K}$.

Likewise, the miss terms for the stage $i=2,\dots,K$ can be given as follows.
\begin{equation}
\begin{aligned}
R_{i,M}^1 &= \int \mathrm{p}(\mathrm{d}y_{1:i}^{1})\Big\{
C_M^1\pi_i^1(y_{1:i}^1)\mathbb{I}(\delta_i^1 = 0,\delta_{i-1}^1 = F^1)\Big\}\\
R_{i,M}^2 &= \int \mathrm{p}(\mathrm{d}y_{1:i}^{1:2})\Big\{
C_M^2\pi_i^2(y_{1:i-1}^{1:2},y_i^2)\mathbb{I}(\delta_i^{2} = 0,\delta_{i-1}^2 = F^2)+\\
& C_M^2\pi_i^2(y_{1:i-1}^{1:2},y_i^1)\mathbb{I}(\delta_i^{2} = 0,\delta_{i-1}^1 = F^1,\delta_{i-1}^2 = F^1)\Big\}\\
\end{aligned}
\end{equation}
where the first term of $R_{i,M}^2$ is again due to using the secondary feature $y_i^2$ ($\delta_{i-1}^2 = F^2$), and the second term is due to using the shared feature $y_i^1$ ($\delta_{i-1}^2 = F^1$ and $\delta_{i-1}^1 = F^1$).
Similarly, the false-alarm (false positive) term at the last stage is given as follows.
\begin{equation}
\begin{aligned}
R_{K,A}^1 = &\int \mathrm{p}(\mathrm{d}y_{1:K}^{1})\Big\{
C_{A}^1(1-\pi_K^1(y_{1:K}^1))\\
&\mathbb{I}(\delta_K^1 = 1,\delta_{K-1}^1 = F^1)\Big\}\\
R_{K,A}^2 = &\int \mathrm{p}(\mathrm{d}y_{1:K}^{1:2})\Big\{
C_{A}^2(1-\pi_K^2(y_{1:K-1}^{1:2},y_K^2))\\
&\mathbb{I}(\delta_K^2 = 1,\delta_{K-1}^2 = F^2)+\\
&C_{A}^2(1-\pi_K^2(y_{1:K-1}^{1:2},y_K^1))\\
&\mathbb{I}(\delta_K^{2} = 1,\delta_{K-1}^1 = F^1,\delta_{K-1}^2 = F^1)\Big\}\\
\end{aligned}
\end{equation}

An important step in solving Problem \eqref{eqn:minR} is the following
expansion of the expected resource cost in \eqref{eqn:lambdaE}.
By the law of total probability,
\begin{equation}\label{eqn:totalProb}
\begin{aligned}
D_1^1 = D_1^1 \Big\{ \mathrm{P}(\delta_1^1=0) + \sum_{i=2}^{K-1} \mathrm{P}(\delta_{i}^1=0,\delta_{i-1}^1=F^1)+\\
\mathrm{P}(\delta_K^1=0,\delta_{K-1}^1=F^1)+\mathrm{P}(\delta_K^1=1,\delta_{K-1}^1=F^1) \Big\}\\
\end{aligned}
\end{equation}
and
\begin{equation}
\begin{aligned}
D_{i+1}^1 \mathrm{P}(\delta_{i}^1=F^1) = D_{i+1}^1 \Big\{ \sum_{j=i+1}^{K-1} \mathrm{P}(\delta_j^1=0,\delta_{j-1}^1=F^1)+\\
\mathrm{P}(\delta_K^1=0,\delta_{K-1}^1=F^1)+\mathrm{P}(\delta_K^1=1,\delta_{K-1}^1=F^1) \Big\},\\
i = 1,\dots,K-1
\end{aligned}
\end{equation}

Similarly for the secondary application
\begin{equation}
\begin{aligned}
D_{i+1}^2 \mathrm{P}(\delta_{i}^2=F^2)= D_{i+1}^2 \Big\{ \sum_{j=i+1}^{K-1} \mathrm{P}(\delta_j^2=0,\delta_{j-1}^2=F^2)+\\
\mathrm{P}(\delta_K^2=0,\delta_{K-1}^2=F^2) + \mathrm{P}(\delta_K^2=1,\delta_{K-1}^2=F^2) \Big\},\\
i = 0,\dots,K-1
\end{aligned}
\end{equation}

Putting everything back into \eqref{eqn:minR} yields a dynamic programming structure, with the state variable being the posteriors $\pi_i^j,j=1,2$ defined in Section \ref{subsec:featModel}. Minimizing \eqref{eqn:minR} can thus be achieved efficiently using the following backward procedure. 
\begin{equation}\label{eqn:VMultiApp}
\begin{aligned}
V_K^1(\pi_K^1) &\triangleq \min_{\delta_K^1}\mathbb{I}(\delta_K^{1} = 0)C_M^1\pi_K^1 + \mathbb{I}(\delta_K^{1} = 1)C_{A}^1(1-\pi_K^1)\\
V_K^2(\pi_K^2) &\triangleq \min_{\delta_K^2}\mathbb{I}(\delta_K^{2} = 0)C_M^2\pi_K^2 + \mathbb{I}(\delta_K^{2} = 1)C_A^2(1-\pi_K^2)\\
V_i^1(\pi_i^1) &\triangleq \min_{\delta_i^1}\mathbb{I}(\delta_i^{1} = 0)C_M^1\pi_i^1 +\\
& \mathbb{I}(\delta_i^{1} = F^1) \Big\{\lambda D_{i+1}^1+\mathbb{E}[V_{i+1}^1(\pi_{i+1}^1(Y_{i+1}^1,\pi_i^1))]\Big\}\\
V_i^2(\pi_i^2;\pi_i^1) &\triangleq \min_{\delta_i^2}\mathbb{I}(\delta_i^{2} = 0)C_M^2\pi_i^2 +\\
& \mathbb{I}(\delta_i^{1\ast} = F^1,\delta_i^{2} = F^1) \mathbb{E}[V_{i+1}^2(\pi_{i+1}^2(Y_{i+1}^1,\pi_i^2);\pi_i^1)] + \\
& \mathbb{I}(\delta_i^{2} = F^2) \Big\{ \lambda D_{i+1}^2+\mathbb{E}[V_{i+1}^2(\pi_{i+1}^2(Y_{i+1}^2,\pi_i^2);\pi_i^1)] \Big\}\\
&i = 1,\dots,K-1\\
V_0^1(\pi_0^1) &\triangleq \lambda D_{1}^1+\mathbb{E}[V_{1}^1(\pi_1^1(Y_{1}^1,\pi_0^1))]\\
V_0^2(\pi_0^2;\pi_0^1) &\triangleq \min_{\delta_0^2}
\mathbb{I}(\delta_0^{2} = F^1) \mathbb{E}[V_{1}^2(\pi_1^2(Y_{1}^1,\pi_0^2);\pi_0^1)] + \\
& \mathbb{I}(\delta_0^{2} = F^2) \Big\{ \lambda D_{1}^2+\mathbb{E}[V_{1}^2(\pi_1^2(Y_{1}^2,\pi_0^2);\pi_0^1)] \Big\}\\
\end{aligned}
\end{equation}
where the expectation is taken with respect to the evidence probabilities (see Section \ref{subsec:featModel})
and $V_i^j$ is the value function at stage $i$ of application $j$. From the first and third expressions of \eqref{eqn:VMultiApp}, the minimizers for the primary application can be obtained by setting
\begin{equation}\label{eqn:deltaOpt1}
\delta_K^{1\ast}(\pi_K^1) = 
\begin{cases}
0, \pi_K^1 <C_A^1/(C_A^1+C_M^1)\\
1, \text{ else}
\end{cases}
\end{equation}
and
\begin{equation}\label{eqn:deltaOpt2}
\begin{aligned}
\delta_i^{1\ast}(\pi_i^1) &= 
\begin{cases}
0, V_i^1(\pi_i^1) = C_M^1\pi_i^1\\
F^1, V_i^1(\pi_i^1) < C_M^1\pi_i^1
\end{cases},\\
&i = 1,\dots,K-1
\end{aligned}
\end{equation}
The expression in \eqref{eqn:deltaOpt2} can be further simplified into
\eqref{eqn:optDecRules} using Lemmas \ref{lem:EVconcave} and \ref{lem:EV0}.

From the second and fourth expressions of \eqref{eqn:VMultiApp}, the optimal decision rule for the secondary application is
\begin{equation}
\delta_K^{2\ast}(\pi_K^2) = 
\begin{cases}
0, \pi_K^2 <C_A^2/(C_A^2+C_M^2)\\
1, \text{ else}
\end{cases}
\end{equation}
and
\begin{equation}\label{eqn:deltaOpt3}
\begin{aligned}
\delta_i^{2\ast}(\pi_i^2;\pi_i^1) &= 
\begin{cases}
0, V_i^2 = C_M^2\pi_i^2,\\
F^2, V_i^2 = \lambda D_{i+1}^2 + \mathbb{E}[V_{i+1}^2(Y_{i+1}^2,\pi_i^2;\pi_i^1)]\\
F^1, V_i^2 = \mathbb{E}[V_{i+1}^2(Y_{i+1}^1,\pi_i^2;\pi_i^1)], \pi_i^1 \geq \tau_i^{1\ast}\\
\end{cases},\\
&i = 0,\dots,K-1
\end{aligned}
\end{equation}
The expression in \eqref{eqn:deltaOpt3} can be further simplified into
\eqref{eqn:optDecRules2} using Lemma \ref{lem:sharing}.

\begin{lemma}\label{lem:EVconcave}
	$\mathbb{E}[V_{i+1}(\pi_{i+1}(Y_{i+1}, \pi))],i = 0,\dots,K-1$ and $V_i(\pi),i=1,\dots,K$ are concave\footnote{Moreover, $V_i(\pi),i=1,\dots,K$ can be shown to be piece-wise linear and concave, which was first observed and proven (by induction) in \cite[Smallwood and Sondik]{smallwood1973optimal}.}.
	\begin{proof}
		$V_K(\pi)$ is concave. Hence, by Lemma \ref{lem:concave}, $\mathbb{E}[V_{K}(\pi_{K}(Y_{K}, \pi))]$ is concave.
		
		Assume that $V_{i+1}(\pi)$ is concave, thus $\mathbb{E}[V_{i+1}(\pi_{i+1}(Y_{i+1}, \pi))]$ is concave by Lemma \ref{lem:concave}, then
		\begin{equation}
		V_i(\pi) = \min \{(\pi),\lambda D_{i+1}+\mathbb{E}[V_{i+1}(\pi_{i+1}(Y_{i+1}, \pi))]
		\end{equation} 
		is also concave. Again, by Lemma \ref{lem:concave}, $\mathbb{E}[V_{i}(\pi_{i}(Y_{i}, \pi))]$ is concave.
	\end{proof}
\end{lemma}

\begin{lemma}\label{lem:concave}
	$\mathbb{E}[V_{i+1}(\pi_{i+1}(Y_{i+1}, \pi))]$ is concave if $V_{i+1}(\pi)$ is concave.
	\begin{proof}
		See \cite[p. 146]{bertsekas1976dynamic}.
	\end{proof}
\end{lemma}

\begin{lemma}\label{lem:EV0}
	$\mathbb{E}[V_{i+1}(\pi_{i+1}(Y_{i+1}, 0))] = 0, i = 0,\dots,K-1$.
	\begin{proof}
		$V_{K}(0) = 0$, then $\mathbb{E}[V_{K}(\pi_{K}(Y_{K}, 0))] = V_{K}(0) = 0$
		
		Assume that $\mathbb{E}[V_{i+1}(\pi_{i+1}(Y_{i+1}, 0))] = 0$, then
		\begin{equation}
		V_i(0) = \min \{0,\lambda D_{i+1}\} = 0. 
		\end{equation}
		Hence, $\mathbb{E}[V_{i}(\pi_{i}(Y_{i}, 0))] = V_{i}(0) = 0$.
	\end{proof}
\end{lemma}

\begin{lemma}\label{lem:sharing}
	If the condition in \eqref{eqn:condForShare} holds, then
	\begin{equation}\label{eqn:42}
	\begin{aligned}
	\delta_i^{2\ast}(\pi_i^2;\pi_i^1) &= 
	\begin{cases}
	0, V_i^2 = \pi_i^2, \pi_i^1 < \tau_i^{1\ast}\\
	F^2, V_i^2 < \pi_i^2, \pi_i^1 < \tau_i^{1\ast}\\
	0, V_i^2 = \pi_i^2, \pi_i^1 \geq \tau_i^{1\ast}\\
	F^1, V_i^2 < \pi_i^2, \pi_i^1 \geq \tau_i^{1\ast}\\
	\end{cases},\\
	&i = 0,\dots,K-1
	\end{aligned}
	\end{equation}
	which implies $\delta_i^{2\ast} \neq F^2$ when $\pi_i^{1} \geq \tau_i^{1\ast}$.
	
	\begin{proof}
		The fourth expression of \eqref{eqn:VMultiApp} is equivalent to Eq.\ \eqref{eqn:42} if and only if
		\begin{equation}\label{eqn:conditions}
		\mathbb{E}[V_{i}^2(Y_{i}^1,\pi_{i-1}^2)]- \mathbb{E}[V_{i}^2(Y_{i}^2,\pi_{i-1}^2)] \leq \lambda D_{i}^2\\
		\end{equation}
		
		The condition in \eqref{eqn:conditions} is made satisfied by \eqref{eqn:condForShare} because
		of Lemma \ref{lem:Ediffconcave} (note that $V_i^2$ is concave over $\pi_i^2$ for each $\pi_i^1$,$i=1,\dots,K$).
	\end{proof}
\end{lemma}

\begin{lemma}\label{lem:Ediffconcave}
	\begin{equation}\label{eqn:bound1,chap:apx_sharing}
	\begin{aligned}
	\mathbb{E}[V_i(Y_i^1,\pi_{i-1})]- \mathbb{E}[V_i(Y_i^2,\pi_{i-1})] &\leq
	C_M^2 \left\{ \mathbb{E}[\pi_i(Y^1_i)] - \mathbb{E}[\pi_i(Y^2_i)]  \right\}\\
	&i = 1,\dots,K
	\end{aligned}
	\end{equation}
	
	\begin{proof}
		Since $V_i(\pi_i)$ is concave, $V_i'(\pi_i)$ is non-increasing. Furthermore, $V_i'(\epsilon) = C_M^2$ for some small $\epsilon > 0$. Therefore, $V_i'(\pi_i) \leq C_M^2$, i.e.\
		\begin{equation}\label{eqn:bound2,chap:apx_sharing}
		V_i(\pi_i(Y_i^1)) - V_i(\pi_i(Y_i^2)) \leq 
		C_M^2 \big[ \pi_i(Y_i^1)-\pi_i(Y_i^2) \big]
		\end{equation}
		
		Taking expectation on both size of \eqref{eqn:bound2,chap:apx_sharing} yields \eqref{eqn:bound1,chap:apx_sharing}.
	\end{proof}
\end{lemma}

\subsection{Proof of Proposition \ref{prop:vivaCascade}}\label{subsec:app2}
Introducing (additional) early positive decisions to intermediate stages results in the following modification to the third and fourth lines of \eqref{eqn:VMultiApp}.
\begin{equation}
\begin{aligned}
V_i^1(\pi_i^1) &\triangleq \min_{\delta_i^1}\mathbb{I}(\delta_i^{1} = 0)C_M^1\pi_i^1 +\mathbb{I}(\delta_i^{1} = 1)C_A^1(1-\pi_i^1)\\
& \mathbb{I}(\delta_i^{1} = F^1) \Big\{\lambda D_{i+1}^1+\mathbb{E}[V_{i+1}^1(\pi_{i+1}^1(Y_{i+1}^1,\pi_i^1))]\Big\}\\
V_i^2(\pi_i^2;\pi_i^1) &\triangleq \min_{\delta_i^2}\mathbb{I}(\delta_i^{2} = 0)C_M^2\pi_i^2 + \mathbb{I}(\delta_i^{2} = 1)C_A^2(1-\pi_i^2)\\
& \mathbb{I}(\delta_i^{1\ast} = F^1,\delta_i^{2} = F^1) \mathbb{E}[V_{i+1}^2(\pi_{i+1}^2(Y_{i+1}^1,\pi_i^2);\pi_i^1)] + \\
& \mathbb{I}(\delta_i^{2} = F^2) \Big\{ \lambda D_{i+1}^2+\mathbb{E}[V_{i+1}^2(\pi_{i+1}^2(Y_{i+1}^2,\pi_i^2);\pi_i^1)] \Big\}\\
&i = 1,\dots,K-1\\
\end{aligned}
\end{equation}
Therefore the positive decision is \textit{not} chosen by the optimal policy under the following circumstances.
\begin{equation}\label{eqn:51}
\begin{aligned}
\delta_i^{j\ast} \neq 1 \text{ if } V_i^j < C_A^j(1-\pi_i^j),\\
i=1,\dots,K-1,j=1,2
\end{aligned}
\end{equation}
Since $V_i^1$ and $V_i^2$ are concave functions of $\pi_i^1$ and $\pi_i^2$, respectively, \eqref{eqn:51} is equivalent to
\begin{equation}
\begin{aligned}
\delta_i^{j\ast} \neq 1 \text{ if } \pi_i^j \leq \max\{\pi_i^j:V_i^j < C_A^j(1-\pi_i^j)\}, \\
i=1,\dots,K-1,j=1,2
\end{aligned}
\end{equation}

Hence if \eqref{eqn:condVivaCascade} holds then the positive decisions are never chosen by the optimal policy, and therefore do not make any difference in the end system performance.

\begin{algorithm}[t]
	\caption{Pseudo-code to find optimal thresholds for the multi-application cascade system. This algorithm has the time complexity of $O(KM^2L)$ and the space complexity of $\max\{O(KM^2),O(M^2L)\}$, where $L$ is the quantization level of the feature models.}\label{alg:recur_thresh}
	\begin{algorithmic}[1]
		\Function{optimize}{$\textit{model}^1$, $\textit{model}^2$}
		\State $\textit{model}^1$,$\textit{model}^2$ are structures containing the primary and secondary application's feature models, respectively
		\State $K$ is the number of stages
		\State $M$ is the state probability quantization size
		\State Use \eqref{eqn:leastFavor} 
		to obtain robust versions of $\textit{model}^1$ and $\textit{model}^2$.
		\State $b=[0:1/(M-1):1]$ (dummy) probability vector 
		\State $V_K^1 = \min(C_{M}^1b, C_{A}^1(1-b))$
		\State $\tau_K^{1\ast}$ = $C_{A}^1/(C_{A}^1+C_{M}^1)$
		\For {$i = 1:1:M$}
		\State $V_K^2(:,i) = \min(C_{M}^2b, C_{A}^2(1-b))$
		\State $\tau_K^{2\ast}(i)$ = $C_{A}^2/(C_{A}^2+C_{M}^2)$
		\EndFor 
		\For {$i = K-1:-1:1$}
		\State $J^1$ = expected next-stage ($i$+1) primary value function
		\State $V_i^1 = \min(C_{M}^1b, J^1)$
		\State $\tau_i^{1\ast} = \min \{b: V_{i}^1 - C_{M}^1b < 0\}$ 
		\State $\tau_i^{1\ast} = \max(\pi_{Li}^1,\min(\pi_{Ui}^1,\tau_i^{1\ast}))$
		\State $J^{2_1}$ = expected next-stage ($i$+1) secondary value function using the shared primary feature
		\State $J^{2_2}$ = expected next-stage ($i$+1) secondary value function using the secondary feature
		\For {$j = 1:1:M$}
		\If {$b(j)< \tau_i^{1\ast}$}
		\State $V_i^2(:,j) = \min (C_M^2b,J^{2_2}(:,j))$
		\State $\tau_i^{2\ast}(j) = \min \{b: V_{i}^2(:,j) - C_{M}^2b < 0\}$
		\State $\tau_i^{2\ast}(j) = \max(\pi_{Li}^2,\min(\pi_{Ui}^2,\tau_i^{2\ast}(j)))$
		\Else
		\State $V_i^2(:,j) = \min (C_M^2b,J^{2_1}(:,j))$
		\State $\eta_i^{2\ast}(j) = \min \{b: V_{i}^2(:,j) - C_{M}^2b < 0\}$
		\State $\eta_i^{2\ast}(j) = \max(\pi_{Li}^2,\min(\pi_{Ui}^2,\eta_i^{2\ast}(j)))$
		\EndIf
		\EndFor 
		\EndFor
		\State $V_0^1 = J^1$ = expected next-stage (1) primary value function
		\State $V_0^2 = J^{2_1}$ = expected next-stage (1) secondary value function using the shared primary feature
		\EndFunction
		%
	\end{algorithmic}
\end{algorithm}

\bibliographystyle{IEEEtran}
\bibliography{main}

\begin{thebibliography}{10}
\providecommand{\url}[1]{#1}
\csname url@samestyle\endcsname
\providecommand{\newblock}{\relax}
\providecommand{\bibinfo}[2]{#2}
\providecommand{\BIBentrySTDinterwordspacing}{\spaceskip=0pt\relax}
\providecommand{\BIBentryALTinterwordstretchfactor}{4}
\providecommand{\BIBentryALTinterwordspacing}{\spaceskip=\fontdimen2\font plus
\BIBentryALTinterwordstretchfactor\fontdimen3\font minus
  \fontdimen4\font\relax}
\providecommand{\BIBforeignlanguage}[2]{{%
\expandafter\ifx\csname l@#1\endcsname\relax
\typeout{** WARNING: IEEEtran.bst: No hyphenation pattern has been}%
\typeout{** loaded for the language `#1'. Using the pattern for}%
\typeout{** the default language instead.}%
\else
\language=\csname l@#1\endcsname
\fi
#2}}
\providecommand{\BIBdecl}{\relax}
\BIBdecl

\bibitem{lee2012terraswarm}
E.~A. Lee, J.~D. Kubiatowicz, J.~M. Rabaey, A.~L. Sangiovanni-Vincentelli,
  S.~A. Seshia, J.~Wawrzynek, D.~Blaauw, P.~Dutta, K.~Fu, C.~Guestrin
  \emph{et~al.}, ``{The TerraSwarm Research Center (TSRC)(A white paper)},''
  \emph{EECS Department, University of California, Berkeley, Tech. Rep.
  UCB/EECS-2012-207}, 2012.

\bibitem{atzori2010internet}
L.~Atzori, A.~Iera, and G.~Morabito, ``{The Internet of Things: A survey},''
  \emph{Computer networks}, vol.~54, no.~15, pp. 2787--2805, 2010.

\bibitem{le2013energy}
L.~Le, D.~M. Jun, and D.~L. Jones, ``Energy-efficient detection system in
  time-varying signal and noise power,'' in \emph{IEEE International Conference
  on Acoustics, Speech and Signal Processing (ICASSP), 2013}.\hskip 1em plus
  0.5em minus 0.4em\relax IEEE, 2013, pp. 2736--2740.

\bibitem{viola2001rapid}
P.~Viola and M.~Jones, ``Rapid object detection using a boosted cascade of
  simple features,'' in \emph{Proceedings of the 2001 IEEE Computer Society
  Conference on Computer Vision and Pattern Recognition, 2001}, vol.~1.\hskip
  1em plus 0.5em minus 0.4em\relax IEEE, 2001, pp. I--511.

\bibitem{salamon2014dataset}
J.~Salamon, C.~Jacoby, and J.~P. Bello, ``A dataset and taxonomy for urban
  sound research,'' in \emph{Proceedings of the ACM International Conference on
  Multimedia}.\hskip 1em plus 0.5em minus 0.4em\relax ACM, 2014, pp.
  1041--1044.

\bibitem{mcaulay1986speech}
R.~J. McAulay and T.~F. Quatieri, ``Speech analysis/synthesis based on a
  sinusoidal representation,'' \emph{IEEE Transactions on Acoustics, Speech and
  Signal Processing}, vol.~34, no.~4, pp. 744--754, 1986.

\bibitem{turaga2006resource}
D.~S. Turaga, O.~Verscheure, U.~V. Chaudhari, and L.~D. Amini, ``Resource
  management for networked classifiers in distributed stream mining systems,''
  in \emph{Sixth International Conference on Data Mining, 2006}.\hskip 1em plus
  0.5em minus 0.4em\relax IEEE, 2006, pp. 1102--1107.

\bibitem{tang1991optimization}
Z.-B. Tang, K.~R. Pattipati, and D.~L. Kleinman, ``Optimization of detection
  networks: {Part I - Tandem structures},'' \emph{IEEE Transactions on Systems,
  Man and Cybernetics}, vol.~21, no.~5, pp. 1044--1059, 1991.

\bibitem{swaszek1993performance}
P.~F. Swaszek, ``On the performance of serial networks in distributed
  detection,'' \emph{IEEE Transactions on Aerospace and Electronic Systems},
  vol.~29, no.~1, pp. 254--260, 1993.

\bibitem{viswanathan1988optimal}
R.~Viswanathan, S.~C. Thomopoulos, and R.~Tumuluri, ``Optimal serial
  distributed decision fusion,'' \emph{IEEE Transactions on Aerospace and
  Electronic Systems}, vol.~24, no.~4, pp. 366--376, 1988.

\bibitem{luo2005optimization}
H.~Luo, ``Optimization design of cascaded classifiers,'' in \emph{IEEE Computer
  Society Conference on Computer Vision and Pattern Recognition, 2005},
  vol.~1.\hskip 1em plus 0.5em minus 0.4em\relax IEEE, 2005, pp. 480--485.

\bibitem{jun2010energy}
D.~M. Jun and D.~L. Jones, ``An energy-aware framework for cascaded detection
  algorithms,'' in \emph{2010 IEEE Workshop on Signal Processing Systems
  (SIPS)}.\hskip 1em plus 0.5em minus 0.4em\relax IEEE, 2010, pp. 1--6.

\bibitem{jun2013cascading}
------, ``Cascading signal-model complexity for energy-aware detection,''
  \emph{IEEE Journal on Emerging and Selected Topics in Circuits and Systems},
  vol.~3, no.~1, pp. 65--74, 2013.

\bibitem{chen2012fidelity}
J.~Chen, R.~Tan, G.~Xing, X.~Wang, and X.~Fu, ``Fidelity-aware utilization
  control for cyber-physical surveillance systems,'' \emph{IEEE Transactions on
  Parallel and Distributed Systems}, vol.~23, no.~9, pp. 1739--1751, 2012.

\bibitem{cohen2013managing}
D.~Cohen, ``Managing resources on a multi-modal sensing device for energy-aware
  state estimation,'' Master's thesis, 2013.

\bibitem{raykar2010designing}
V.~C. Raykar, B.~Krishnapuram, and S.~Yu, ``Designing efficient cascaded
  classifiers: tradeoff between accuracy and cost,'' in \emph{Proceedings of
  the 16th ACM SIGKDD International Conference on Knowledge Discovery and Data
  Mining}.\hskip 1em plus 0.5em minus 0.4em\relax ACM, 2010, pp. 853--860.

\bibitem{chen2012classifier}
M.~Chen, K.~Q. Weinberger, O.~Chapelle, D.~Kedem, and Z.~Xu, ``Classifier
  cascade for minimizing feature evaluation cost,'' in \emph{International
  Conference on Artificial Intelligence and Statistics}, 2012, pp. 218--226.

\bibitem{emre1999polarimetric}
B.~Emre~Ertin, ``Polarimetric processing and sequential detection for automatic
  target recognition systems,'' Ph.D. dissertation, The Ohio State University,
  1999.

\bibitem{trapeznikov2013multi}
K.~Trapeznikov, V.~Saligrama, and D.~Casta{\~n}{\'o}n, ``Multi-stage classifier
  design,'' \emph{Machine learning}, vol.~92, no. 2-3, pp. 479--502, 2013.

\bibitem{trapeznikov2013supervised}
K.~Trapeznikov and V.~Saligrama, ``Supervised sequential classification under
  budget constraints,'' in \emph{Proceedings of the Sixteenth International
  Conference on Artificial Intelligence and Statistics}, 2013, pp. 581--589.

\bibitem{wang2014lp}
J.~Wang, K.~Trapeznikov, and V.~Saligrama, ``An {LP} for sequential learning
  under budgets.'' in \emph{AISTATS}, 2014, pp. 987--995.

\bibitem{jun2013cheap}
D.~M. Jun, L.~Le, and D.~L. Jones, ``Cheap noisy sensors can improve activity
  monitoring under stringent energy constraints,'' in \emph{Global Conference
  on Signal and Information Processing (GlobalSIP), 2013 IEEE}.\hskip 1em plus
  0.5em minus 0.4em\relax IEEE, 2013, pp. 683--686.

\bibitem{huber1968robust}
P.~J. Huber, ``Robust confidence limits,'' \emph{Zeitschrift f{\"u}r
  Wahrscheinlichkeitstheorie und verwandte Gebiete}, vol.~10, no.~4, pp.
  269--278, 1968.

\bibitem{huber2011robust}
------, ``Robust statistics,'' \emph{International Encyclopedia of Statistical
  Science}, pp. 1248--1251, 2011.

\bibitem{levy2008principles}
B.~C. Levy, \emph{Principles of signal detection and parameter
  estimation}.\hskip 1em plus 0.5em minus 0.4em\relax Springer, 2008.

\bibitem{debruin15powerblade}
\BIBentryALTinterwordspacing
S.~DeBruin, B.~Ghena, Y.-S. Kuo, and P.~Dutta, ``Powerblade: A low-profile,
  true-power, plug-through energy meter,'' in \emph{Proceedings of the 13th ACM
  Conference on Embedded Networked Sensor Systems}, ser. SenSys '15.\hskip 1em
  plus 0.5em minus 0.4em\relax New York, NY, USA: ACM, 2015. [Online].
  Available: \url{http://doi.acm.org/10.1145/2809695.2809716}
\BIBentrySTDinterwordspacing

\bibitem{leonard2010variation}
W.~J. Leonard, J.~Neal, and R.~Ratnam, ``Variation of {Type B} song in the
  endangered {Golden-cheeked Warbler} ({Dendroica} chrysoparia),'' \emph{The
  Wilson Journal of Ornithology}, vol. 122, no.~4, pp. 777--780, 2010.

\bibitem{chamberland2004asymptotic}
J.-F. Chamberland and V.~V. Veeravalli, ``Asymptotic results for decentralized
  detection in power constrained wireless sensor networks,'' \emph{IEEE Journal
  on Selected Areas in Communications}, vol.~22, no.~6, pp. 1007--1015, 2004.

\bibitem{smallwood1973optimal}
R.~D. Smallwood and E.~J. Sondik, ``The optimal control of partially observable
  {Markov} processes over a finite horizon,'' \emph{Operations Research},
  vol.~21, no.~5, pp. 1071--1088, 1973.

\bibitem{bertsekas1976dynamic}
D.~P. Bertsekas, ``Dynamic programming and stochastic control,'' 1976.

\end{thebibliography}
%
%
%

%

\balance
\begin{IEEEbiography}[{\includegraphics[width=1in,height=1.25in,clip,keepaspectratio]{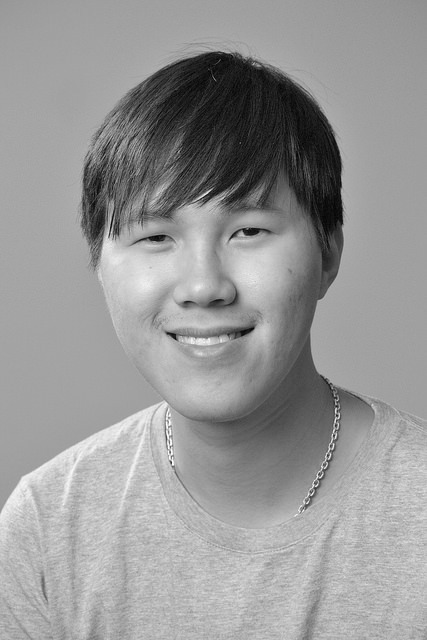}}]{Long N. Le}
	received the BSEE and MSEE degrees in
	electrical engineering in 2011 and 2013 from the Ho Chi Minh University of Technology and the University of Illinois at Urbana-Champaign, respectively.
	During the summer of 2015, he was an intern at the Audio and Acoustics Research Group at Microsoft Research. He was awarded the Best-in-Session by the Semiconductor Research Corporation at TECHCON 2016, in the IoT System Design session.
	He is currently working toward the Ph.D. degree as a research assistant at the
	Coordinated Science Laboratory and Beckman Institute of the University of Illinois at Urbana-Champaign. 
	His current research interests include resource-efficient statistical inference and signal processing, with a focus on IoT applications.
\end{IEEEbiography}
\begin{IEEEbiography}[{\includegraphics[width=1in,height=1.25in,clip,keepaspectratio]{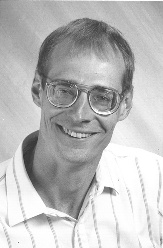}}]{Douglas L. Jones}
	received the BSEE, MSEE,
	and Ph.D. degrees from Rice University, Houston,
	TX, USA, in 1983, 1986, and 1987, respectively.
	During the 1987-1988 academic year, he was at
	the University of Erlangen-Nuremberg, Germany,
	on a Fulbright postdoctoral fellowship. Since 1988,
	he has been with the University of Illinois, Urbana-Champaign, IL, USA, where he is currently the director of Advanced Digital Sciences Center (ADSC) and a 
	Professor in Electrical and Computer Engineering,
	Neuroscience, the Coordinated Science Laboratory,
	and the Beckman Institute. He was on sabbatical
	leave at the University of Washington in Spring 1995 and at the University of
	California at Berkeley in Spring 2002. In the Spring semester of 1999 he served
	as the Texas Instruments Visiting Professor at Rice University. His research
	interests are in digital signal processing and systems, including nonstationary
	signal analysis, adaptive processing, multisensor data processing, OFDM,
	and various applications such as low-power implementations, biology and
	neuroengineering, and advanced hearing aids and other audio systems. He is
	an author of two DSP laboratory textbooks.
	Dr. Jones served on the Board of Governors of the IEEE Signal Processing
	Society from 2002 to 2004. He was selected as the 2003 Connexions Author of
	the Year.
\end{IEEEbiography}






\end{document}